\documentclass[seceq]{ptptex}

\usepackage[dvips]{graphicx}
\usepackage{wrapft}
\input epsf

\title{  
 Generalized S-matrix in Mixed Representation }

\author{
Kenzo  \textsc{Ishikawa} and Takashi \textsc{Shimomura }}

\inst{
Department of Physics, Hokkaido University, 
             Sapporo 060-0810, Japan
}

\abst{
A generalized scattering amplitude where momenta of  incoming-particles  
and outgoing-particles as well  as positions  of   incoming-particles  and 
outgoing-particles are specified is formulated. Idealistic  beams and  
idealistic measuring instruments where  momenta and positions satisfy 
minimum uncertainty are studied with a use of minimum wave packets, 
coherent states.  In the present work, we show general features of the
 generalized scattering amplitudes based on ${\phi}^4$ theory. We give 
 a proof of completeness of many body states, asymptotic behaviors in the large
 distance region, and factorization of the amplitudes. Despite of the 
non-orthogonal properties of wave packets, we found that the probability 
interpretation is verified.  A differential probability depends upon
 the wave packet size but a total probability that is integrated in 
the final states is independent from the size of final state wave packet
and becomes universal. Few body amplitudes are studied as examples. 
}

\begin{document}
\maketitle

\section{Introduction}
We study scattering processes in which both of momenta and positions are 
simultaneously measured. Observations of both variables are made in
recent neutrino experiments where a distance between a source region and
a detector is fixed in a macroscopic length  and the momenta of the
particles are
measured. Scattering amplitudes of both variables are formulated in the
present paper.
   
In
quantum mechanics, precise 
values of momentum and coordinate of a particle can not be measured 
simultaneously from Heisenberg uncertainty relation.  So in scattering 
processes, one of the variables is selected
and a dependence of the scattering amplitude on this variable is
studied.  Normally  momentum variable is selected and
the momentum dependence of  the transition 
probability is studied and is compared  with a theoretical
calculation. Since  momenta are commuting variables, it is possible to 
determine them precisely. In real experiments, however, the exact value of
momentum is difficult to measure and the momentum is measured with finite 
uncertainty due  to a finite resolution of a detector. When the
 momentum is measured with uncertainties,
simultaneous measurements of positions and  momenta
become   possible in experiments. The positions are measured
 also within finite uncertainties  due  to spatial resolution of detectors. 
When momenta and positions are measured with finite
uncertainties, total number of information could  be 
equivalent with the standard case where the exact values of momenta are 
measured.  

Since a wave function with a definite momentum is a plane
wave and  is invariant under any translations, any particular position
is undefined. But a wave
function with a finite uncertainty of the momentum can have a finite
spatial extension and is localized around certain position, so in this
case the spatial position is defined. So it is 
necessary to introduce finite uncertainty of momenta in order to define
the positions. As a price of a finite
uncertainty on the  momentum, it became possible to introduce a position 
where the particle is measured or is produced. The position has also
finite uncertainty. 
In this work, we introduce such amplitude that is defined 
with finite uncertainties of momenta and positions. 

The wave functions with finite spatial extensions are necessary actually
for asymptotic conditions of scattering processes to be 
satisfied \cite{Goldberger}\cite{Sasakawa}. However in many situations of
high energy physics, effects of finite wave 
functions have been ignorable and it has been sufficient to use
propagators of  momentum variables  with  $i\epsilon $ prescription. 
We clarify these points and  we study situations where the 
effects of finite wave functions are
important. To investigate these problems we define the generalized
S-matrix of both variables using mixed representation and we show
several features of the 
scattering amplitudes in which the momentum and the positions are measured. Some
implications and applications are also studied.

We formulate a position-dependent and momentum-dependent S-matrix
in an idealistic limit allowed from Heisenberg uncertainty relation. 
In our formalism, each momentum and position of the initial particle and 
final particle satisfy minimum uncertainty relation. This corresponds
to a scattering process where coordinates as well as momenta of the beam 
satisfy the minimum uncertainty relation and the measurement are made
with minimum uncertainty allowed from Heisenberg uncertainty relation.

In real experiments the uncertainty of the momentum and 
coordinate may be larger than  those of the present work. But our
idealistic scattering matrix should be realized in a suitable method 
and should give new  insights in quantum mechanical scattering processes.   
Also extensions to non-minimum wave packets is straightforward.

Uncertainties of momenta and positions  are actually 
determined in experiments by resolutions of beam sources and of
detectors. The resolutions depend on each accelerator and detector. So
we leave the uncertainties unfixed and study the scattering matrix of
this situation.

Because the momentum is conjugate to the coordinate, they satisfy 
the commutation relation,
\begin{equation}
[x,p]=i,
\end{equation}
in a unit ${h/ 2\pi} =1$ and a momentum resolution  and a spatial resolution  of a detector satisfy 
Heisenberg's uncertainty relation,
\begin{equation}
\delta p \times \delta x \ge  { 1  \over 2}.
\end{equation}

A product of uncertainties  becomes  minimum in coherent  states. The
coherent state of a variable, $x$, 
\begin{eqnarray}
& &\langle x |P_0,X_0\rangle =N_1e^{i{ P_0 } (x-X_0)-{1 \over 2{\sigma}}(x-X_0)^2} \\
& &N_1^2=(\pi {\sigma})^{-{1 \over 2}}
\end{eqnarray} 
has expectation values
\begin{eqnarray}
& &\langle| x| \rangle = X_0      \\
& &\langle| x^2| \rangle= {X_0}^2+{1 \over 2}{\sigma}      \\
& &\langle| p| \rangle= P_0         \\
& &\langle| p^2| \rangle={P_0}^2+{1 \over {2 {\sigma}}}. 
\end{eqnarray}
The product between the variances of the momentum and coordinate,   
\begin{eqnarray}
& &(\delta x)^2 \times (\delta p)^2= {1 \over 4} \\
& &( \delta x)^2= \langle| x^2| \rangle-(\langle| x| \rangle)^2 \\
& &( \delta p)^2= \langle| p^2| \rangle-(\langle| p| \rangle)^2 
\end{eqnarray} 
is independent from $\sigma $ and satisfies the minimum uncertainty 
condition. The uncertainty becomes minimum 
with the coherent state.   

The coherent state satisfies also completeness condition,
\begin{eqnarray}
& &\int {d P_0 d X_0 \over 2 \pi }\langle x| P_0,X_0\rangle \langle  P_0,X_0|y\rangle \nonumber \\
& &= \int {d P_0 d X_0 \over 2 \pi } N_1^2 e^{iP_0(x-y)}e^{-{1 \over 2 \sigma }(x-X_0)^2-{1 \over 2 \sigma }(y-X_0)^2 }\\
& &=\delta (x-y). \nonumber
\end{eqnarray}

We use these wave functions for  expressing in-states and out-states. In the
former, the state corresponds to the idealistic beam and in the latter the state
corresponds to the idealistic measuring apparatus. A transition
element between the states of idealistic beams and the states of  the
idealistic measurement is computed from  these  matrix elements. 
Using  them, we define the idealistic scattering matrix in mixed 
representation and study its properties and applications.

Our formalism will be applied to long line experiments such as,
solar neutrino experiments, atmospheric  neutrino experiments, long base
line neutrino experiments, reactor neutrino experiments, and others  where 
systems have huge scales and positions of detectors play important
roles \cite{Kayser},\cite{Giunti},\cite{Cardall},\cite{Beuthe}\cite{Asahara} \cite{Ishikawa-shimomura}. In addition to neutrino, scattering of other weakly 
interacting particles where position dependence in addition to momentum 
dependence are measured and give important physical informations will be
studied.      

 The paper is organized in the following manner. In Section 2,
 mathematics on wave packets are given. Wave packet wave functions are
 explicitly given and the overlap integrals and time dependent
 behaviors are analyzed. New uncertainty relations 
between the velocity
 of expansions in asymptotic region and the initial sizes
are obtained.  Although some of the materials in this section  may be
known to experts, they are necessary and useful for later arguments.  
In Section 3, generalized scattering amplitude is defined and general 
properties are studied. We find suitable integration measures for both
variables in  which the  probability interpretation is verified.  A 
differential probability depends upon the wave packet size but a total 
probability that is integrated in the final states is independent from
the size of final state wave packet and becomes universal. Explicit 
examples are given in Section 4 and summary is given in Section 5.      
\section{Wave packets }
\subsection{Mathematical preliminaries}
\subsubsection{Complete sets of minimum wave packets :time independent wave packets}
Let  momentum eigenstates and position eigenstates be $|{\vec p}
\rangle$ and  $|{\vec x} \rangle$. Transformation from a  momentum 
eigenstate to a position eigenstate is made   from 
\begin{eqnarray}
& &\langle {\vec x}|{\vec p} \rangle=  ({2\pi})^{-{3 \over 2}}e^{i {\vec p} \cdot {\vec x}},  \\
& &\langle {\vec x}|{\vec p} \rangle^*=\langle {\vec p}|{\vec x} \rangle=\langle {\vec x}|{\vec -p} \rangle.
\end{eqnarray}
Both sets of functions are  complete sets and satisfy 
\begin{eqnarray}
& &\int d{\vec x} \langle {\vec p}_1|{\vec x} \rangle \langle {\vec x}|{\vec p}_2 \rangle =\delta({\vec p}_1-{\vec p}_2), \\
& &\int d{\vec p} \langle {\vec x}_1|{\vec p} \rangle \langle {\vec p}|{\vec x}_2 \rangle =\delta({\vec x}_1-{\vec x}_2), 
\end{eqnarray}
and 
\begin{eqnarray}
& &\int d{\vec x} |{\vec x} \rangle \langle {\vec x}| = 1,  \\
& &\int d{\vec p} |{\vec p} \rangle \langle {\vec p}| = 1. 
\end{eqnarray}

The normalized coherent states in three spatial dimensions \cite{completeness} are  defined
as
\begin{eqnarray}
\label{eq:coherentx}
& &\langle {\vec x} |{\vec P}_0,{\vec X}_0\rangle =N_3e^{i{ {\vec P}_0 } ({\vec x}-{\vec X}_0)-{1 \over 2{\sigma}}({\vec x}-{\vec X}_0)^2} \\
& &N_3^2=(\pi {\sigma})^{-{3 \over 2}},\nonumber  \\
& &\langle {\vec x} |{\vec P}_0,{\vec X}_0\rangle^* = \langle {\vec P}_0,{\vec X}_0|{\vec x} \rangle= \langle {\vec x} |-{\vec P}_0,{\vec X}_0\rangle.
\end{eqnarray} 
Due to the Gaussian factor the wave function is localized around the
center position ${\vec x}_0$ with a width $\sqrt{2 \sigma}$ and is an 
approximate eigenfunction of the momentum operator.
A constant $N_3$ is determined from the normalization condition
\begin{eqnarray}
& &\int d^3 x |\langle {\vec x} |{\vec P}_0,{\vec X}_0\rangle|^2\nonumber \\
& &=N_3^2 \int d^3 x  e^{-{1 \over {\sigma}}({\vec x}-{\vec X}_0)^2} \\  
& &=N_3^2(\pi \sigma)^{{3 \over 2}} \nonumber\\
& &=1.\nonumber
\end{eqnarray}
 
The same states are expressed in the momentum representation as 
\begin{eqnarray}
\label{eq:coherentp}
& & \langle {\vec p}|{\vec P}_0,{\vec X}_0 \rangle = \int d {\vec x} 
\langle {\vec p}|{\vec x} \rangle \langle {\vec x}|{\vec P}_0,{\vec X}_0 
\rangle \nonumber\\ 
& &=N_3{\sigma}^{3/2} e^{-i{{\vec p} \cdot {\vec X}_0} -{\sigma \over 2}
({\vec P}_0-{\vec p})^2},   \\                   
& &\langle {\vec p} |{\vec P}_0,{\vec X}_0\rangle^* = \langle {\vec P}_0,{\vec X}_0|{\vec p} \rangle= \langle {\vec p} |-{\vec P}_0,{\vec X}_0\rangle
\end{eqnarray}
and satisfy the normalization condition in the momentum representation.

The set of functions for an arbitrary value of $\sigma$ satisfy the
completeness condition in  the coordinates representation,
\begin{equation}
\int{ {d{\vec P}_0}{d{\vec X}_0} \over (2 \pi )^3}\langle{\vec x}_1 |{\vec P}_0,{\vec X}_0\rangle \langle {\vec P}_0,{\vec X}_0|{\vec x}_2\rangle=\delta({\vec x}_1-{\vec x}_2),
\end{equation}
and the completeness condition in  the momentum  representation,
\begin{equation}
\int{ {d{\vec P}_0}{d{\vec X}_0} \over (2 \pi)^3}\langle{\vec p}_1 |{\vec P}_0,{\vec X}_0\rangle \langle {\vec P}_0,{\vec X}_0|{\vec p}_2 \rangle=\delta({\vec p}_1-{\vec p}_2).
\end{equation}
The completeness condition is simply expressed as 
\begin{equation}
\label{eq:coherent-complete}
\int{ {d{\vec P}_0}{d{\vec X}_0} \over (2 \pi )^3}|{\vec P}_0,{\vec X}_0\rangle \langle {\vec P}_0,{\vec X}_0|= 1.
\end{equation}

Overlap between two coherent states are computed from  the above
definition
\begin{eqnarray}
& &\langle {\vec P}_1,{\vec X}_1| {\vec P}_2,{\vec X}_2 \rangle \nonumber\\
&= &\int d^3 x\langle {\vec P}_1,{\vec X}_1|{\vec x}\rangle \langle{\vec x}| {\vec P}_2,{\vec X}_2 \rangle \nonumber \\
&=&N_3^2 \int d^3 x e^{i{(P_1-P_2) \over \hbar}x+ i({P_1 \over \hbar }X_1-{P_2 } X_2)- {1 \over 2\sigma^2}\left(({\vec x}-{\vec X}_1)^2+({\vec x}-{\vec X}_2)^2\right)} \\
&=&e^{-{1 \over 4\sigma}({\vec X}_1-{\vec X}_2)^2-{\sigma \over 4}({\vec P}_1-{\vec P}_2)^2+ {i \over 2}({\vec P}_1+{\vec P}_2)({\vec X}_1-{\vec X}_2)}  \nonumber .
\end{eqnarray}

Obviously matrix elements do not vanish and states are not orthogonal
for different values of the momenta and coordinates. Despite of the
nonorthogonality they satisfy
\begin{equation}
\int{ {d{\vec P}_0}{d{\vec X}_0} \over (2 \pi )^3}|{\vec P}_0,{\vec X}_0\rangle \langle {\vec P}_0,{\vec X}_0| {\vec P},{\vec X} \rangle  =  |{\vec P},{\vec X} \rangle .
\end{equation}
Hence $\langle {\vec P}_1,{\vec X}_1| {\vec P}_2,{\vec X}_2 \rangle $
plays a role of the Dirac delta function.

\subsubsection{Wave packets defined at arbitrary time}

Wave packets defined at a certain time $T_0$ is constructed when one particle
energy is known. Let  $E({\vec p})$ stands one particle energy of the
momentum ${\vec p}$, 
\begin{eqnarray}
E({\vec p})=({\vec p}^2+m^2)^{1/2}
\end{eqnarray}
in the unit with $c=1$, then the wave packet defined at  $T_0$ is,
\begin{equation}
\label{eq:coherentp-time}
\langle {\vec p}|{\vec P}_0,{\vec X}_0,T_0 \rangle=\langle {\vec p}|{\vec P}_0,{\vec X}_0\rangle e^{{-E({\vec p}) \over i } T_0}.
\end{equation}
This set of functions for a given time $T_0$ satisfy the completeness condition,
\begin{equation}
\int{ {d{\vec P}_0}{d{\vec X}_0} \over (2 \pi )^3}\langle {\vec p}_1 |{\vec P}_0,{\vec X}_0
,T_0 \rangle \langle {\vec P}_0,{\vec X}_0,T_0|{\vec  p}_2 \rangle =\langle {\vec p}_1|{\vec p}_2 \rangle.
\end{equation}
The representation of these wave packets in the coordinates space is
obtained by the Fourier transformation,
\begin{equation}
\label{eq:coherentx-time}
\langle {\vec x}|{\vec P}_0,{\vec X}_0,T_0 \rangle= \int d^3 p \langle {\vec x}|{\vec p} \rangle \langle {\vec p}|{\vec P}_0,{\vec X}_0\rangle e^{{-E({\vec p}) \over i } T_0}.
\end{equation}  
It is easy  to verify the completeness condition by combining
Eq.(\ref{eq:coherent-complete}) and (\ref{eq:coherentp-time}) with Eq.(\ref{eq:coherentx-time}),
\begin{equation}
\int{ {d{\vec P}_0}{d{\vec X}_0} \over (2 \pi )^3} \langle {\vec x}_1|{\vec P}_0,{\vec X}_0
,T_0 \rangle \langle {\vec P}_0,{\vec X}_0,T_0| {\vec x}_2 \rangle =\langle {\vec x}_1|{\vec x}_2 \rangle. 
\end{equation}
The completeness condition is simply written as 
\begin{equation}
\label{eq:coherent-tcomplete}
\int{ {d{\vec P}_0}{d{\vec X}_0} \over (2 \pi )^3} |{\vec P}_0,{\vec X}_0
,T_0 \rangle \langle {\vec P}_0,{\vec X}_0,T_0| =1. 
\end{equation}
The matrix elements of the wave packets at different times are computed as,
\begin{equation}
\langle {\vec P}_1,{\vec X}_1,T_1|{\vec P}_2,{\vec X}_2
,T_2 \rangle=\int d^3 p\langle {\vec P}_1,{\vec X}_1,T_1|{\vec p}\rangle \langle 
{\vec p}|{\vec P}_2,{\vec X}_2,T_2 \rangle
\end{equation}
and satisfy
\begin{eqnarray}
& &\int{ {d{\vec P}_0}{d{\vec X}_0} \over (2 \pi )^3} |{\vec P}_0,{\vec X}_0
,T_0 \rangle \langle {\vec P}_0,{\vec X}_0,T_0| {\vec P}_1,{\vec X}_1,T_1 \rangle = |{\vec P}_1,{\vec X}_1,T_1\rangle, \\
& &\langle {\vec P}_1,{\vec X}_1,T_1| {\vec P}_2,{\vec X}_2,T_1 \rangle =\langle {\vec P}_1,{\vec X}_1| {\vec P}_2,{\vec X}_2 \rangle. 
\end{eqnarray}
Explicit forms of these matrix elements are given later.
We use these wave functions for  expanding the field operator.

\subsection{Matrix elements}

\subsubsection{Time dependent transformation function}
We calculate the matrix elements of the mixed states appeared in the
previous section. It is convenient to define time dependent
transformation functions, 
\begin{eqnarray}
& &\langle t,{\vec p}|{\vec P}_0,{\vec X}_0,T_0 \rangle \nonumber \\
& &=e^{{E({\vec p}) \over i} (t-T_0) }\langle {\vec p}|{\vec P}_0,{\vec X}_0 \rangle \\
& &=N_3{\sigma}^{3/2}e^{{E({\vec p}) \over i}(t-T_0) }e^{-i{\vec p}\cdot {\vec X}_0-{\sigma \over 2}({\vec p}-{\vec P}_0)^2}\nonumber  
\end{eqnarray}
and   
\begin{eqnarray}
& &\langle t,{\vec x}|{\vec P}_0,{\vec X}_0,T_0 \rangle \nonumber \\
& & = \int d{\vec p} \langle {\vec x}|{\vec p}\rangle \langle t,{\vec p}|{\vec P}_0,{\vec X}_0,T_0 \rangle\\
& &=N_3 ({\sigma \over 2\pi})^{3/2} \int d{\vec p} e^{{E({\vec p}) \over i}(t-T_0)+ i{\vec p}\cdot{\vec x}-
i{\vec p} \cdot {\vec X}_0-{\sigma \over 2}({\vec p}-{\vec P}_0)^2} \nonumber\\
 & &= N_3 ({\sigma \over 2\pi})^{3/2} \int d{\vec p} e^{-iE({\vec p})(t-T_0)+ i{\vec p}\cdot( {\vec x}-{\vec X}_0)-{\sigma \over 2}({\vec p}-{\vec P}_0)^2}. \nonumber
\end{eqnarray}
The absolute value of the integrand becomes maximum at ${\vec P}_0$ but
the phase becomes large in large $t-T_0$ region or large ${\vec x}-{\vec
X}_0$ region. So we integrate on ${\vec p}$ in two regions
separately. In the small $t-T_0$ region we use the approximation of the
integrand around the  ${\vec P}_0$  and in the large $t-T_0$ region we
use the approximation of the integrand around the stationary momentum.

(A) Small $T-T_0$ case:translational motion.

In the small $t-T_0$ region, the integral is written
and computed around ${\vec P}_0$ as
\begin{eqnarray}
\label{eq:closetime}
& &\langle t,{\vec x}|{\vec P}_0,{\vec X}_0,T_0 \rangle \nonumber \\
& &= N_3 ({\sigma \over 2\pi})^{3/2} \int d{\vec p} e^{-i\left(E({\vec p}_0)+({\vec p}-{\vec p}_0) \cdot {\vec v}_0\right)(t-T_0)+ i\left ({\vec p}_0+({\vec p}-{\vec p}_0)\right)\cdot( {\vec x}-{\vec X}_0)-{\sigma \over 2}({\vec p}-{\vec P}_0)^2} \nonumber\\
& &=N e^{i \phi}, \\
& &N=N_3 e^{-{1 \over 2 \sigma}({\vec x}-{\vec X}_0-{\vec v}_0(t-T_0))^2 },
\\
& &e^{i\phi}=e^{-iE({\vec P}_0)(t-T_0)+i{\vec P}_0 \cdot ({\vec x}-{\vec X}_0)}, \\
& &{\vec v}_0={\partial \over \partial p_i}E({\vec p})|_{{\vec p}={\vec p}_0}.
\end{eqnarray}  
The wave packet keeps its shape and moves with a constant velocity
${\vec v}_0$. The center of wave packet is 
${\vec X}_0$ at $t=T_0$ and is   ${\vec X}_0+{\vec v}_0(t-T_0)$ at a
time $t$. 

(B) Large $t-T_0$: expanding wave packet.

In the large $t-T_0$ region the momentum integration applied in the
previous method is not a good approximation any more  because the phase
oscillates rapidly in this region. The phase of the integrand  becomes
stationary at ${\vec P}_X$ which satisfies 
\begin{equation}
{\partial \over \partial p_i} \left({-iE({\vec p})(t-T_0)+ i{\vec p}\cdot( {\vec x}-{\vec X}_0)-{\sigma \over 2}({\vec p}-{\vec P}_0)^2} \right)=0.
\end{equation}
The solution is obtained by expanding the momentum  in ${1 \over t-T_0}$
and is given by, 
\begin{eqnarray}
& &{\vec P}_X={\vec P^{(0)}}_X+{\vec P^{(1)}}_X+{\vec P^{(2)}}_X, \\
& &{\vec P^{(0)}}_X= m{1 \over \sqrt{(t-T_0)^2-({\vec x}-{\vec X}_0)^2 } }({\vec x}-{\vec X}_0),  \\
& &{\vec P^{(1)}}_X= i{1 \over t-T_0}  \sigma E({\vec P}_X)({\vec P}_X-{\vec P}_0),   \\
& &{\vec P^{(2)}}_X=o((t-T_0)^{-2}). 
\end{eqnarray}
So the exponent of the integrand is expanded around ${\vec P}_X$ and we 
have   
\begin{eqnarray}
\label{eq:asymptotic}
& &\langle t,{\vec x}|{\vec P}_0,{\vec X}_0,T_0 \rangle \\
& &= N e^{-iE({\vec p})(t-T_0)+ i{\vec p}\cdot( {\vec x}-{\vec X}_0)  }, \nonumber \\
& &N=N_3({1 \over 2i{ \gamma_L \over \sigma}+1})^{1/2}({1 \over 2i{ \gamma_T \over \sigma}+1})  e^{-{1 \over 2 }\sigma ({{\vec P^{(0)}}_X}-{{\vec P}_0})^2 +\delta},  \\
& &\delta=   {1 \over 2 }\sigma ({{\vec P}_X}-{{\vec P}_0})^2( 2i-{\xi} )\xi,
 \\
& &\xi={ \sigma E({\vec P}_X) \over (t-T_0) }, 
\end{eqnarray}   
in a large $t-T_0$ where the wave packet parameters are given by  
\begin{eqnarray}
& &\gamma_L={1 \over 2}{m^2 |t-T_0| \over E({\vec P}_X)^3}, \\
& &\gamma_T={1 \over 2}{|t-T_| \over E({\vec P}_X)}.
\end{eqnarray}
The longitudinal component and transverse component of a momentum ${\vec
q}$ are defined as 
\begin{eqnarray} 
\label{eq:decomposition}
& &{\vec q}_T={\vec q}-{\vec P}_X{({\vec P}_X,{\vec q}) \over ({\vec P}_X,{\vec P}_X)}               \\
& &{\vec q}_L={\vec P}_X{({\vec P}_X,{\vec q}) \over ({\vec P}_X,{\vec P}_X)}.
\end{eqnarray}
The phase factor in the Eq.(\ref{eq:asymptotic}) is written in leading
order of $|t-T_0|$ as 
\begin{eqnarray}
& &e^{ {-i(t-T_0)E({\vec P}_X)+i{\vec P}_X \cdot ({\vec x}-{\vec X}_0)})}\nonumber
\\
& &=e^{-i{ m \over } \sqrt{(t-T_0)^2-({\vec x}-{\vec X}_0)^2} } \\
& &=e^{{-i{m^2 \over  E({\vec P}_X) }(t-T_0)}}.\nonumber
\end{eqnarray}
This phase factor becomes very small if the mass is very small and
vanishes in the massless case. The fact that the phase becomes small in
the high energy region or in the small mass region is a characteristic
property of relativistic invariant theory\cite{classical-phase}. 

The absolute magnitude of $N$ becomes maximum when the momentum 
${\vec P}_X$ agrees to ${\vec P}_0$. This is realized at a particular
position of ${\vec x}$ , ${\vec x}_0$, which satisfies    
\begin{equation}
{\vec P}_0= m{1 \over \sqrt{(t-T_0)^2-({\vec x}_0-{\vec X}_0)^2 } }({\vec x}_0-{\vec X}_0).   
\end{equation}
The solution is 
\begin{equation}
{\vec x}_0={\vec X}_0+(t-T_0){  {\vec P}_0\over E({\vec P}_0) }, 
\end{equation}
where ${\vec x}_0$ is the center of wave packet at a time $t$. 

We write the difference of the momenta, $ {\vec P}_X-{\vec P}_0$, in the 
Gaussian exponent in Eq.$(\ref{eq:asymptotic})$ using ${\vec x}$ and 
${\vec x}_0$ and a unit vector ${\vec n}_1$ in the ${\vec p}_1$ direction  as,
\begin{equation}
{\vec P}_X-{\vec P}_0={E({\vec P}_0) \over t-T_0}\{({\vec x}-{\vec x}_0 )_T+
{E({\vec P}_0)^2 \over m^2}|({\vec x}-{\vec x}_0)_L|{\vec n}_1 \}.
\end{equation}
We substitute this expression into Eq.$(\ref{eq:asymptotic})$  and we have  
\begin{eqnarray}
& &\exp({-{1 \over 2}\sigma ({\vec P}_X-{\vec P}_0)^2 }) \\
& &=\exp({ -{1 \over 2}\sigma  {(E({\vec P}_0))^2 \over (t-T_0)^2}{({\vec x}-{\vec x}_0 )_T}^2 -{1 \over 2}\sigma {(E({\vec P}_0))^6 \over m^4 (t-T_0)^2}{({\vec x}-{\vec x}_0)_L}^2}). \nonumber
\end{eqnarray}
The normalization factor in  the small $|t-T_0|$ region has also  
\begin{equation}
\exp({-{1 \over 2 \sigma} {({ {\vec x}-{\vec x}_0})}^2}).
\end{equation}
Hence the size of the wave packet in the longitudinal direction  is
given as  
\begin{equation}
\label{longitudinal-size}
{\delta x}_L=\sqrt{ {2 \over \sigma}} {m^2 |t-T_0| \over E({\vec P}_0)^3}+\sqrt{2 \sigma}
\end{equation}
and in the transverse direction is given as  
\begin{equation}
\label{transverse-size}
{\delta x}_T=\sqrt{{ 2\over \sigma}} { |t-T_0| \over E({\vec P}_0)}+\sqrt{2 \sigma}.
\end{equation}
A wave packet expansion is characterized by its velocity. The velocity of 
expansion in the transverse direction, $v_T$, is determined by the momentum 
variance as
\begin{equation}
{ v}_T=\sqrt{{ 2\over \sigma}} { 1 \over E({\vec P}_0)}.
\end{equation}
and that in the longitudinal direction,$v_L$, is determined as    
\begin{equation}
{ v}_L=\sqrt{{ 2\over \sigma}} { m^2 \over (E({\vec P}_0))^3}.
\end{equation}

The velocity of expansion satisfy uncertainty relations 
\begin{eqnarray}
\label{new-uncertainty1}
& &v_T \delta x(t=0) E({\vec P}_0)=1 \\
\label{new-uncertainty2}
& &v_L \delta x(t=0) E({\vec P}_0)=({ m \over E({\vec P}_0)})^2,
\end{eqnarray}
where $\delta x(t=0)$ is the spatial extension of wave packet at $t=0$.
The $v_L$ is given by multiplying ${m^2 \over E({\vec P}_0)^2}$ to the 
$v_T$. This ratio becomes one in the non-relativistic energy  region
where    $E({\vec p}_0)$ is nearly equal to $m$. Consequently  wave
packets expand symmetrically in the non-relativistic region.  In the
relativistic region where $E({\vec p}_0)$ is much larger than $m$, the
ratio between both values becomes very small and wave packets expand 
an-symmetrically. The shape becomes a circular thin disk after certain time.  
The wave packet size is given as a function of the propagation
time, t, for various values of the initial wave packet size in
Fig(1). The $ {\delta x}_T$ becomes huge size in 500 sec.,the period
between the sun and the earth.
%
%
\begin{figure}[th]
\centerline{
\epsfbox{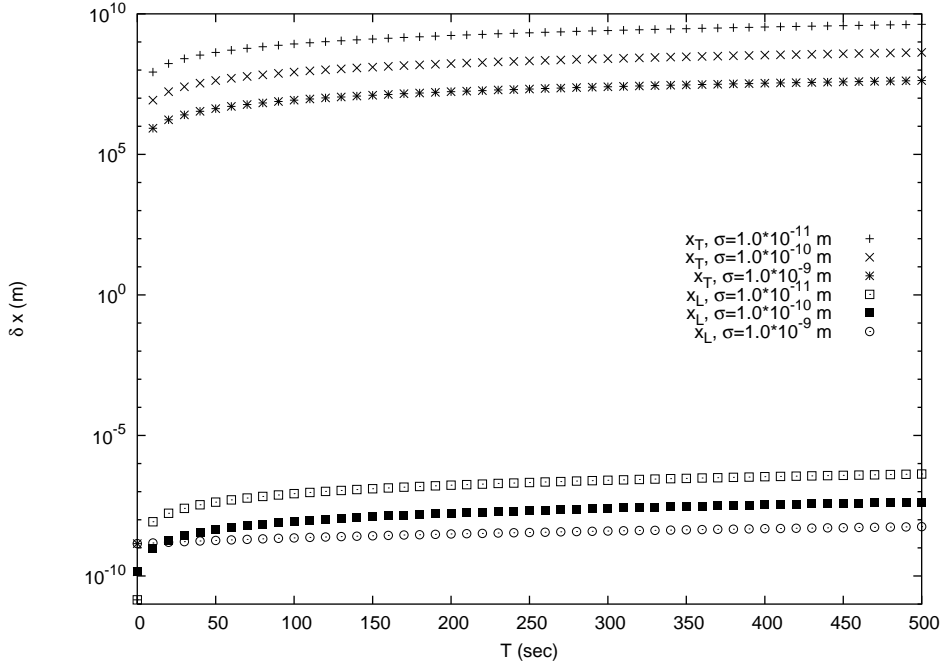}}
\caption{  Time dependence of wave packet size in the physical unit is given.
$\delta_T$  is the size in the transverse direction given in  
Eq.(\ref{transverse-size}) and $\delta_L$ is the size in the
longitudinal direction given in  Eq.(\ref{longitudinal-size}) . 
$mc^2=10^{-2}eV $ is assumed.
}
\label{fig:1}
\end{figure}

Due to the expansion of wave packet, the normalization of the wave
function is inversely proportional to $\gamma_T \sqrt {\gamma_L}$ in the
asymptotic region.

\subsubsection{Two body matrix elements}
Various matrix elements of wave packets are studied in this section.

 1.Matrix elements of wave packets defined at equal  time \\

The overlap between two wave packets is given as,     
\begin{eqnarray}
& &\langle {\vec P}_1,{\vec X}_1|{\vec P}_2,{\vec X}_2\rangle \\
&=&e^{-{1 \over 4\sigma}({\vec X}_1-{\vec X}_2)^2-{\sigma \over 4}({\vec P}_1-{\vec P}_2)^2}e^{{i \over 2}({\vec P}_1+{\vec P}_2)({\vec X}_1-{\vec X}_2)}  \nonumber .
\end{eqnarray}
Thus the overlap decreases fast as the distance between two positions
in the real space and in the momentum become large.  Matrix elements for 
the same coordinates is 
\begin{equation}
\langle {\vec P}_1,{\vec X}|{\vec P}_2,{\vec X} \rangle=e^{-{\sigma \over 4}({\vec P}_1-{\vec P}_2)^2}
\end{equation}
and for the same momenta is 
\begin{equation}
\langle {\vec P},{\vec X}_1|{\vec P},{\vec X}_2 \rangle=e^{-{1 \over 4 \sigma}
({\vec X}_1-{\vec X}_2)^2 + i{{\vec P}_1} \cdot ({\vec X}_1-{\vec X}_2)}.
\end{equation}
From these matrix elements if a particle is prepared at $({\vec P}', {\vec X}')$  probabilities of finding a particle in a region of the 
momentum and coordinate between $({\vec P}, {\vec X})$ and
$({\vec P}+d{\vec P}, {\vec X}+d{\vec X})$ 
is given by 
\begin{equation}
P= {d {\vec P} d {\vec X} \over (2 \pi )^3} e^{-{1 \over 2\sigma}({\vec X}-{\vec X'})^2-{\sigma \over 2}({\vec P}-{\vec P'}_2)^2}
\end{equation}
Heisenberg uncertainty relation between the variance in the coordinates
and the variance in the momenta is satisfied. It is
important to notice that the momentum and the coordinate is measured 
simultaneously. 

2. Time dependent matrix elements \\
The time dependent matrix element 
is computed by inserting the complete set of momentum  eigenstates,
\begin{eqnarray}
& &\langle {\vec P}_1,{\vec X}_1,T_1| {\vec P}_2,{\vec X}_2,T_2 \rangle \nonumber \\
& &=\int d^3 p\langle {\vec P}_1,{\vec X}_1|{\vec p}\rangle e^{E({\vec p})(T_1-T_2) \over {i}}\langle{\vec p}|{\vec P}_2,{\vec X}_2 \rangle  \\
& &= {N_3}^2{\sigma}^3 e^{-{ \sigma \over 4} ({\vec P}_1-{\vec P}_2)^2  }\int d^3 p e^{{E({\vec p})(T_1-T_2) \over {i}}+i{\vec p}\cdot({\vec X}_1-{\vec X}_2) -\sigma({\vec p}-{\vec P}_0)^2},  \nonumber \\
& &{\vec P}_0= {1 \over 2}({\vec P}_1+{\vec P}_2).\nonumber 
\end{eqnarray}
In the small  $|T_1-T_2|$ region this  matrix element is given as,
\begin{eqnarray}
& &\langle {\vec P}_1,{\vec X}_1,T_1| {\vec P}_2,{\vec X}_2,T_2 \rangle\\
& &=N e^{-iE({\vec P}_0)(T_1-T_2)+i{\vec P}_0 \cdot ({\vec X}_1-{\vec X}_2)},\nonumber \\
& &N=e^{ -{1 \over 4\sigma}({\vec X}_1-{\vec X}_2-{\vec v}(T_1-T_2))^2-
{\sigma \over 4}({\vec P}_1-{\vec P}_2)^2 }
\end{eqnarray}
where the velocity is given by
\begin{equation}
{\vec v}_i={\partial E({\vec p}) \over \partial p_i}|_{\vec p}={{\vec P}_0 \over E({\vec P})}.
\end{equation}
This matrix element shows that the shape of the wave packet is preserved
but the center position is moving with the velocity ${\vec v}$ in the
small $|T_1-T_2|$ region.
 
The  matrix element in  a large $|T_1-T_2|$ region is obtained by  the 
stationary phase approximation in ${\vec p}$ integration. The stationary 
point ${\vec P}_X$ is obtained from the stationarity condition,
\begin{equation}
{\partial  \over \partial p_i} (i{\vec p} \cdot ({\vec X}_1-
{\vec X}_2)-iE({\vec p})(T_1-T_2) -\sigma ({\vec p}-{\vec P}_0)^2) =0,
\end{equation} 
and is given as,
 \begin{eqnarray}
& &{\vec P}_X=({\vec X}_1-{\vec X}_2) {m \over ((T_1-T_2)^2-({\vec X}_1-
{\vec X}_2)^2)^{1/2}}+i{1 \over T_1-T_2} \delta {\vec P}_x,\\
& &\delta {\vec P}_X=2 \sigma  E({\vec P}_X)({\vec P}_X-{\vec P}_0).
\end{eqnarray}
Using this momentum, we have
\begin{eqnarray}
& &\langle {\vec P}_1,{\vec X}_1,T_1| {\vec P}_2,{\vec X}_2,T_2 \rangle\nonumber \\
& &={\tilde N}\exp({ i\phi}), \\
& &\tilde N= (1+{i(T_1-T_2) \over 2 \sigma E( {\vec P}_X)})^{-1}
(1+{i(T_1-T_2)m^2 \over 2 \sigma E({\vec P}_X)^3 } )^{-1/2}\\
& &\times e^{ -{\sigma \over 4}\{({\vec P}_1-{\vec P}_2)^2 +4({\vec P}_X-{\vec P}_0)^2 }, \nonumber \\
& & \exp(i{\phi})= \exp({-iE({\vec P}_X)(T_1-T_2)+i{\vec P}_X({\vec X}_1-{\vec X}_2)}) \\
& &=\exp({i m\sqrt {(T_1-T_2)^2-({\vec X}_1-{\vec X}_2)^2}})\nonumber\\
& &=exp({im^2{|T_1-T_2| \over E({\vec P}_X)}}) \nonumber 
\end{eqnarray}
In the above equations, a transverse component and a longitudinal
component of the momentum are defined as Eq.$(\ref{eq:decomposition})$.
The variances in the transverse directions and the 
longitudinal direction,
$\sigma_T$ and $\sigma_L$, are given as  
\begin{eqnarray}
& &\sigma_T={\sigma \over {1+({T_1-T_2 \over 2\sigma E({\vec P}_X)})^2}} \\
& &\sigma_L={\sigma \over {1+({(T_1-T_2) m^2 \over 2\sigma {E({\vec P}_X)}^3})^2}}.
\end{eqnarray}
Thus the variances $\sigma_T$ and $\sigma_L$ decrease with time. The
spatial extension of the wave packet in the transverse direction is 
inversely proportional to ${\sigma_T}^{1/2}$ and the extension of the 
wave packet in the longitudinal direction is inversely proportional to 
${\sigma_L}^{1/2}$. Hence the wave packet expands differently in
two directions and the expansion in the longitudinal direction is
determined by the absolute value of mass. For the massless case or very
small mass case, a wave packet does not expand in the longitudinal
direction, since the velocity is the constant in this case and $\sigma_L$ 
is equal to $\sigma$.

The stationary momentum 
${\vec P}_X$ is proportional to the direction  ${\vec X}_1-{\vec
X}_2$. After the wave packet expands with time, the wave becomes 
almost spherical wave which is a linear combination of many plane
waves. When the measurement is made at the position ${\vec X}_2$, the 
corresponding wave has a wave vector which is proportional to 
${\vec X}_1-{\vec X}_2$ and the phase which is proportional to the
proper time in the asymptotic region.

 In the asymptotic region,$|T_1-T_2| \rightarrow \infty$, the matrix
 element behaves as,
\begin{eqnarray}
\label{eq:asym-amp}
& &\int d^3 p\langle {\vec P}_1,{\vec X}_1|{\vec p}\rangle e^{E({\vec p})(T_1-T_2) \over {i}}\langle{\vec p}|{\vec P}_2,{\vec X}_2 \rangle={\tilde N} 
\exp(i\phi), \\
& &\tilde N= ({i(T_1-T_2) \over 2 \sigma E( {\vec P}_X)})^{-1}
({i(T_1-T_2)m^2 \over 2 \sigma E({\vec P}_X)^3 } )^{-1/2}\exp({-{\sigma \over 2}\{({\vec P}_X-{\vec P}_1)^2 +({\vec P}_X-{\vec P}_2)^2\} }), \nonumber\\
& &     \\
& & \exp({i\phi})= \exp({i m\sqrt {(T_1-T_2)^2-({\vec X}_1-{\vec X}_2)^2}}). 
\end{eqnarray}  
The probability to find a particle with a momentum ${\vec P}_2$  at a position 
${\vec X}_2$ is given as
\begin{equation}
P=d{\vec P}_2 d{\vec X}_2 {1 \over (2 \pi )^3}|\tilde N|^2 \exp({-\sigma \{({\vec P}_X-{\vec P}_1)^2 +({\vec P}_X-{\vec P}_2)^2\}}).
\end{equation}      

 For a massless case or an extremely small mass case, in  large time  of
$|T_1-T_2|$, the matrix  element Eq.( $\ref{eq:asym-amp}$) is replaced
with,
\begin{eqnarray}
& &\int d^3 p\langle {\vec P}_1,{\vec X}_1|{\vec p}\rangle e^{E({\vec p})(T_1-T_2) \over {i}}\langle{\vec p}|{\vec P}_2,{\vec X}_2 \rangle={\tilde N} 
\exp(i\phi), \\
& &\tilde N= ({i(T_1-T_2) \over 2 \sigma E( {\vec P}_X)})^{-1}
\exp({-{\sigma \over 2}\{({\vec P}_X-{\vec P}_1)^2 +({\vec P}_X-{\vec P}_2)^2\} }),    \nonumber    
\end{eqnarray}  

\section{Generalized scattering matrix}
We study many particle systems where one particle wave functions are 
described by wave packets. A field operator is expanded by a complete
set of wave packets and coefficients become operators.

\subsection{Expansion of field}

A field operator $\phi(x)$ is expanded by a set of momentum  eigenstates as 
\begin{equation}
\label{eq:field expansion}
\phi({\vec x},t)=\int d{\vec p}(\langle {\vec p}|{\vec x} \rangle {1 \over \sqrt{2 \omega({\vec p})}}a({\vec p},t)+\langle {\vec x}|{\vec p} \rangle {1 \over \sqrt{2 \omega({\vec p})}} a^{\dagger}({\vec p},t)). 
\end{equation}
Operators $a({\vec p},t)$ and its conjugate $a({\vec p},t)^{\dagger}$
are creation and
annihilation operator of the momentum ${\vec p}$.

The operator $A({\vec P}_0,{\vec X}_0 ,T_0,t)$ and its conjugate are
defined as 
linear combinations of the operators $a({\vec p,t})$ and its conjugate,
\begin{eqnarray}
\label{eq:atoA}
& &A({\vec P}_0,{\vec X}_0,T_0,t )=\int d{\vec p}  a({\vec p},t) \langle {\vec p}|{\vec P}_0,{\vec X}_0,T_0\rangle  \\
& &A^{\dagger}({\vec P}_0,{\vec X}_0,T_0,t )=\int d{\vec p}\langle {\vec P}_0,{\vec X}_0,T_0|{\vec p}\rangle  a^{\dagger}({\vec p},t).
\end{eqnarray}

Conversely $a({\vec p},t)$ and its conjugate are  solved from the above
equations by using Eq.(\ref{eq:coherent-tcomplete} ) as  
\begin{eqnarray}
& &a({\vec p},t)=  \int { d{\vec P}_0 d{\vec X}_0 \over (2 \pi )^3} A({\vec P}_0,{\vec X}_0,T_0,t) \langle {\vec P}_0,{\vec X}_0,T_0|{\vec p}\rangle \\
& &a^{\dagger}({\vec p},t)=  \int { d{\vec P}_0 d{\vec X}_0 \over (2 \pi )^3} \langle {\vec p}|{\vec P}_0,{\vec X}_0,T_0\rangle A^{\dagger}({\vec P}_0,{\vec X}_0,T_0,t)
\end{eqnarray}
The wave packet size $\sigma$ in  in-state is determined from  the beam 
and the $\sigma$ in  out-state is determined from  the detector 
and they are different generally. For simplicity, we use the same value 
and omit to write in most parts of the present paper. In several places 
we write $\sigma$ explicitly.

By substituting the above expansions to Eq.$(\ref{eq:atoA})$, we have 
\begin{eqnarray}
& &A({\vec P}_0,{\vec X}_0,T_0,t)= \int { d{\vec P}_0' d{\vec X}_0' \over (2 \pi )^3} A({\vec P}_0',{\vec X}_0',T_0',t)\langle {\vec P}_0',{\vec X}_0',T_0'|{\vec P}_0,{\vec X}_0,T_0\rangle  \nonumber \\
& & \\
& &A^{\dagger}({\vec P}_0,{\vec X}_0,T_0,t)= \int { d{\vec P}_0' d{\vec X}_0' \over (2 \pi )^3} \langle {\vec P}_0,{\vec X}_0,T_0|{\vec P}_0',{\vec X}_0'
,T_0'\rangle A^{\dagger}({\vec P}_0',{\vec X}_0',T_0',t).\nonumber \\
& & 
\end{eqnarray}
These equations show that in the space of operator 
$A({\vec P}_0,{\vec X}_0,T_0,t)$ the transformation function  
$\langle {\vec P}_0',{\vec X}_0',T_0' |{\vec P}_0,{\vec X}_0,T_0 \rangle$ 
plays a role of Dirac's delta function.

By substituting the above expansion to Eq.(\ref{eq:field expansion}), the
same field operator is expanded by a set of minimum wave packets of the
center momentum, ${\vec P}_0$, the center coordinates, ${\vec X}_0$, and
the time, $T_0$ as,
\begin{eqnarray}
\label{eq:field expansionA}
& &\phi({\vec x},t)= \int { d{\vec P}_0 d{\vec X}_0 \over (2 \pi )^3} (
C({\vec P}_0,{\vec X}_0,T_0;\vec x) A({\vec P}_0,{\vec X}_0,T_0,t )\\
& &+ C({\vec P}_0,{\vec X}_0,T_0;\vec x)^* A^{\dagger}({\vec P}_0,{\vec X}_0,T_0,t )) ,\nonumber \\
& & C({\vec P}_0,{\vec X}_0,T_0;\vec x)=\int {d {\vec p} \over \sqrt{2 \omega ({\vec p})}}\langle {\vec P}_0,{\vec X}_0,T_0|{\vec p}\rangle \langle {\vec p}|{\vec x} \rangle 
\end{eqnarray}
is obtained. It is convenient to define transformation matrices,
\begin{equation}
 \tilde C({\vec P}_0,{\vec X}_0,T_0;\vec x)=\int d {\vec p}  \sqrt{2 \omega ({\vec p})} \langle {\vec P}_0,{\vec X}_0,T_0|{\vec p}\rangle \langle {\vec p}|{\vec x} \rangle.  
\end{equation}
These  matrices  satisfy  
\begin{eqnarray}
\label{eq:Ctransformation}
& &C({\vec P}_0,{\vec X}_0,T_0;\vec x)=\int { d{\vec P}_1 d{\vec X}_1 \over (2 \pi )^3}
\langle {\vec P}_0,{\vec X}_0,T_0|
{\vec P}_1,{\vec X}_1,T_1 \rangle  C({\vec P}_1,{\vec X}_1,T_1;\vec x) \nonumber \\
& & \\
& &\tilde C({\vec P}_0,{\vec X}_0,T_0;\vec x)=\int { d{\vec P}_1 d{\vec X}_1 \over (2 \pi )^3}
\langle {\vec P}_0,{\vec X}_0,T_0|
{\vec P}_1,{\vec X}_1,T_1 \rangle \tilde C({\vec P}_1,{\vec X}_1,T_1;\vec x)\nonumber \\
& & \\
& &\int d{\vec x} C({\vec P}_0,{\vec X}_0,T_0;\vec x)(\tilde C({\vec P}_0,{\vec X}_0,T_0;\vec x))^*=\langle {\vec P}_0,{\vec X}_0,T_0|{\vec P}_1,{\vec X}_1,T_1 \rangle.\nonumber\\
\end{eqnarray}
\subsection{Complete set of many body states}
Many body state is constructed by the second quantized operators and the
vacuum. In the momentum representation 
the operators $a({\vec p})$ and $a^{\dagger}({\vec p})$ satisfy
\begin{eqnarray}
\label{eq:acommutation}
& &[a({\vec p},t),a^{\dagger}({\vec p'},t')]\delta(t-t')  =\delta({\vec p}-{\vec p'})\delta(t-t')\\
\label{eq:avacuum}
& &a({\vec p},t)|0\rangle=0. 
\end{eqnarray}
A complete set of many body states are constructed from 
\begin{equation}
|0\rangle,|{\vec p}_1\rangle,|{\vec p}_1,{\vec p}_2 \rangle,|{\vec p}_1,{\vec p}_2,{\vec p}_3\rangle,-, |{\vec p}_1,{\vec p}_2,-,-{\vec p}_N\rangle, 
\end{equation}
where 
\begin{eqnarray}
& &|{\vec p}_1\rangle =a^{\dagger}({\vec p}_1,t|0 \rangle,   \\
& &|{\vec p}_1,{\vec p}_2 \rangle =a^{\dagger}({\vec p}_1,t)a^{\dagger}({\vec p}_2,t) |0\rangle,\\
& &|{\vec p}_1,{\vec P}_2,-,-{\vec p}_N  \rangle=\Pi_l a^{\dagger}({\vec p}_l,t)|0 \rangle.
\end{eqnarray}
These particle states satisfy orthonormality conditions with Dirac delta 
function,
\begin{eqnarray}
& &\langle {\vec p}_1 |{\vec p}_1'\rangle =\delta({\vec p}_1-{\vec p}_1'),   \\
& &\langle {\vec p}_1,{\vec p}_2|{\vec p}_1',{\vec p}_2' \rangle =\delta({\vec p}_1-{\vec p}_1')\delta({\vec p}_2-{\vec p}_2')+permutation.
\end{eqnarray}

Let define a projection operator $I$ as 
\begin{eqnarray} 
& & I=\nonumber \\
& &|0\rangle \langle 0|+\int d{\vec p}a^{\dagger}({\vec p})|0\rangle \langle 0| a({\vec p})
+\int {d{\vec p}_1 d{\vec p}_2 \over 2!} a^{\dagger}({\vec p}_1)a^{\dagger}({\vec p}_2)|0\rangle \langle 0| a({\vec p}_1) a({\vec p}_2) \nonumber \\
& &+\int {1 \over l!} {\Pi_l {d{\vec p}_l }
a^{\dagger}({\vec p}_l)}|0\rangle \langle 0| {\Pi_l a({\vec p}_l)} + \cdots ,
\end{eqnarray}
and multiply $I$ to a state
\begin{equation}
|\Psi \rangle =\int f({\vec q}_i)\Pi_l a^{\dagger}({\vec p}_l)|0\rangle.
\end{equation} 
Then from  Eq.(\ref{eq:acommutation}) and (\ref{eq:avacuum}), we have 
\begin{eqnarray}
& &I |\Psi \rangle=\int \Pi_l d^3q_l  f({\vec q}_i)\int {1 \over l!} {\Pi_l {d{\vec p}_l }
a^{\dagger}({\vec p}_l)}|0\rangle \delta({\vec q}_l-{\vec p}_j) \nonumber
\\
& &=\int f({\vec q}_i)\Pi_l a^{\dagger}({\vec p}_l)|0\rangle \nonumber \\
& &=|\Psi \rangle 
\end{eqnarray}
Hence the completeness condition,
\begin{equation}
\label{eq:acompleteness}
I=1
\end{equation}
is satisfied.
The time $t$ is arbitrary and is omitted in the above equations.

Similarly the operators in the mixed representation satisfy 
\begin{eqnarray}
\label{eq:Acommutation}
& &[ A( {\vec P}_0,{\vec X}_0,T_0,t ),A^{\dagger}({\vec P}_0',{\vec X}_0 ',T_0',t') ]\delta(t-t')=
\langle {\vec P}_0,{\vec X}_0,T_0|{\vec P}_0',{\vec X}_0',T_0' \rangle\delta(t-t') \nonumber \\
& & \\
& &A( {\vec P}_0,{\vec X}_0,T_0,t )|0\rangle=0.
\label{eq:Avacuum}
\end{eqnarray}
The particle states defined by these operators are normalized and are
not orthogonal even though momenta and positions are different. However  
the  same complete set is constructed by the vacuum and creation
operators in the mixed representation. Let define an projection operator $I'$, 
\begin{eqnarray} 
& &I'=       \nonumber \\
& &|0\rangle \langle 0|+\int {d{\vec P} d{\vec X} \over (2\pi)^3}A^{\dagger}({\vec P},{\vec X},T)|0\rangle \langle 0| A({\vec P},{\vec X},T) \nonumber \\
&+&\int {1\over 2} {d{\vec P}_1 d{\vec X}_1 \over (2\pi )^3} {d{\vec P}_2 d{\vec X}_2 
\over (2\pi)^3}  A^{\dagger}({\vec P}_1,{\vec X}_1,T_1)A^{\dagger}({\vec P}_2,{\vec X}_2,T_2)|0 \rangle \langle 0| A({\vec P}_2,{\vec X}_2,T_2)A({\vec P}_1,{\vec X}_1,T_1)\nonumber \\
&+&\int {1 \over l!}{\Pi_l {d{\vec P}_l d{\vec X}_l \over (2\pi )^3}}{\Pi_l A^{\dagger}({\vec P}_l,{\vec X}_l,T_l)}|0\rangle \langle 0|{\Pi_l A({\vec P}_l,{\vec X}_l,T_l)} + \cdots 
\end{eqnarray}
The time $t$ in the operators is arbitrary and is omitted in the above 
equations. Let multiply the operator $I'$ to an state,
\begin{equation}
|\Psi' \rangle =\int{\Pi_l {d{\vec Q}_l d{\vec Y}_l \over (2\pi )^3}} F({\vec Q}_i,{\vec Y}_i,S_i)\Pi_l A^{\dagger}({\vec Q}_l,{\vec Y}_l,S_i)|0\rangle.
\end{equation} 
Then from  Eq.(\ref{eq:Acommutation}) and (\ref{eq:Avacuum}), we have 
\begin{eqnarray}
& &I' |\Psi' \rangle \nonumber \\
&=&\int {\Pi_l {d{\vec Q}_l d{\vec Y}_l \over (2\pi )^3}} F({\vec Q}_i,{\vec Y}_i,S_i) 
\int {1 \over L!} {\Pi_m {d{\vec P}_m d{\vec X}_m \over (2\pi )^3}
A^{\dagger}({\vec P}_m,{\vec X}_m,T_m)}|0\rangle \nonumber \\
& &\langle 0|A({\vec P}_m,{\vec X}_m,T_m) \Pi_l A^{\dagger}({\vec Q}_l,{\vec Y}_l,S_i)|0\rangle\nonumber \\
&= &\int{\Pi_l {d{\vec Q}_l d{\vec Y}_l \over (2\pi )^3}} F({\vec Q}_i,{\vec Y}_i,S_i){\Pi_m {d{\vec P}_m d{\vec X}_m \over (2\pi )^3}} A^{\dagger}({\vec P}_m,{\vec X}_m,T_m)\langle {\vec P}_m,{\vec X}_m,T_m|{\vec Q}_l,{\vec Y}_l,S_l \rangle  |0\rangle \nonumber \\
&= &\int{\Pi_l {d{\vec Q}_l d{\vec Y}_l \over (2\pi )^3}} F({\vec Q}_i,{\vec Y}_i,S_i) A^{\dagger}({\vec Q}_l,{\vec Y}_l,T_i) |0\rangle \\
&= &|\Psi' \rangle.\nonumber
\end{eqnarray}
Hence the completeness condition,
\begin{equation}
\label{eq:Acompleteness}
I'=1
\end{equation}
is satisfied.
The time $t$ is arbitrary and is omitted in the above equations.

\subsection{Time evolution}
A unitary operator which translates a time of field operators by a
finite value, t, is given by a Hamiltonian $H$ as,
\begin{equation}
U(t)=e^{{H t \over i}}.
\end{equation}
In a free scalar theory the Hamiltonian  is given as,
\begin{equation}
H=\int d^3 x( {1 \over 2}(\pi(x)^2  +{1 \over 2}({\vec \nabla}\phi(x))^2  +
{1 \over 2}m^2{\phi(x)}^2).
\end{equation}
A commutation relation between the field operator and its conjugate, 
\begin{equation}
\label{eq:field-commutation}
[\phi(x),\pi(y)]\delta(x_0-y_0)=i \delta^{(4)}(x-y)
\end{equation} 
leads the operators $a({\vec p})$ and $a^{\dagger}({\vec p})$ in
Eq.(\ref{eq:field expansion})
satisfy the equal time commutation relation Eq.(\ref{eq:acommutation}).
The Hamiltonian is expressed as
\begin{eqnarray}
& &H=\int d^3p E({\vec p})(a^{\dagger}({\vec p})a({\vec p})+{1 \over 2}) \\
& &E({\vec p})=\sqrt{{\vec p}^2+m^2}\nonumber
\end{eqnarray}
and satisfy 
\begin{eqnarray}
& &[H,a^{\dagger}({\vec p})]=E({\vec p})a^{\dagger}({\vec p})\\
& &[H,a({\vec p})]=-E({\vec p})a^{\dagger}({\vec p})
\end{eqnarray}
From the above commutation relations, we have the time dependence of the 
creation and annihilation operators in the momentum space, 
\begin{eqnarray}
& &a^{\dagger}({\vec p},t) \nonumber \\
&=&U(t)a^{\dagger}({\vec p},0) U^{\dagger}(t)=e^{{E({\vec p})t \over i }}a^{\dagger}({\vec p},0) \\
& &a({\vec p},t)\nonumber \\
&= &U(t)a({\vec p},0) U^{\dagger}(t)=e^{-{E({\vec p})t \over i }}a({\vec p},0).
\end{eqnarray}
The creation and annihilation  operators in the momentum space change
only the c-number phase with time. The states created by the creation
operators in the momentum space stay in the same state,
\begin{eqnarray}
& &U(t)|\Psi \rangle=e^{{\sum_j E({\vec p}_j) \over i}t}|\Psi \rangle \\
& &|\Psi \rangle=a^{\dagger}({\vec p}_1,0)a^{\dagger}({\vec p}_2,0)----a^{\dagger}({\vec p}_l,0)|0\rangle.\nonumber
\end{eqnarray}

  The creation operators and annihilation operators in  the mixed space 
satisfy a commutation relation Eq.(\ref{eq:Acommutation}).
The overlap function in the right hand side is computed in the next 
subsection. This function does not vanish even though the momentum or
the coordinates are  different. Hence the  operator  of  one  set of 
center coordinate and center momentum do not commute with the operator
of a different coordinate and a different momentum, but a set of
operators  satisfy completeness condition from Eq.(\ref{eq:Acompleteness}). 
The operators in the mixed space evolve with time as,  
\begin{eqnarray}
& &A^{\dagger}({\vec P}_0,{\vec X}_0,T_0,t)\\
&=&U(t)A^{\dagger}({\vec P}_0,{\vec X}_0,T_0,o) U^{\dagger}(t)=\int d^3p e^{{E({\vec p})t \over i }}\langle {\vec P}_0,{\vec X}_0,T_0|{\vec p}\rangle a^{\dagger}({\vec p},0) \nonumber\\
&=& \int {d{\vec P}_0'd{\vec X}_0' \over (2\pi )^3 }\int d^3p e^{{E({\vec p})t \over i }}\langle {\vec P}_0,{\vec X}_0,T_0| {\vec p}\rangle \langle {\vec p}|{\vec P}_0',{\vec X}_0',T_0\rangle A^{\dagger}({\vec P'}_0,{\vec X'}_0,T_0,0 )    \nonumber \\
& &A({\vec P}_0,{\vec X}_0,T_0,t)\\
&=& U(t)A({\vec P}_0,{\vec X}_0,T_0,0) U^{\dagger}(t)=\int d^3p e^{-{E({\vec p})t \over i }}\langle {\vec p}| {\vec P}_0,{\vec X}_0,T_0\rangle a({\vec p},t) \nonumber \\
&=&\int {d{\vec P}_0'd{\vec X}_0' \over (2\pi )^3 }\int d^3p e^{-{E({\vec p})t \over i }}\langle {\vec P}_0',{\vec X}_0',T_0| {\vec p}\rangle \langle {\vec p}|{\vec P}_0,{\vec X}_0,T_0\rangle A({\vec P'}_0,{\vec X'}_0,T_0,0 ).\nonumber 
\end{eqnarray}
The time dependent phase factor is not factored out and 
the states created  by the operators  in the mixes space,$A^{\dagger}({\vec P}_0,{\vec X}_0)$ change with time,
\begin{eqnarray}
& &U(t)|\Psi' \rangle \ne e^{{\sum_j E({\vec p}_j) \over i}t}|\Psi' \rangle 
\\
& &|\Psi' \rangle=A^{\dagger}({\vec P}_1,{\vec X}_1)A^{\dagger}({\vec P}_2,{\vec X}_2)----A^{\dagger}({\vec P}_l,{\vec X}_l)|0\rangle .\nonumber
\end{eqnarray}
The matrix elements in the above equations are obtained in  the next
subsection and  we will see that these states are approximate eigenstates.

\subsection{Generalized scattering amplitude and transition probability}
A scattering amplitude where particles in the initial  state of
momentum ${\vec P^i}_l$ are prepared at positions ${\vec X^i}_l$ and times 
${T^i_l}$ and particles in the final state of momentum ${\vec P^o}_m$ are 
measured at positions ${\vec X^o}_m$ and  times ${T^o_m}$ are described as,
\begin{eqnarray}
& &S_{out,in}  \\
&= &\langle {\vec P^o}_1{\vec X^o}_1{ T^o}_1;-;{\vec P^o}_L{\vec X^o}_L{ T^o}_L;out |in ;{\vec P^i}_1{\vec X^i}_1{ T^i}_1;-;{\vec P^i}_M{\vec X^i}_M{ T^i}_M\rangle, \nonumber \\
& &|in ;{\vec P^i}_1{\vec X^i}_1{ T^i}_1;-;{\vec P^i}_M{\vec X^i}_M{ T^i}_M\rangle \\
&=&A^{\dagger}({\vec P^i}_1,{\vec X^i}_1,T^i_1)A^{\dagger}({\vec P^i}_2,{\vec X^i}_2,T^i_2)--A^{\dagger}({\vec P^i}_l,{\vec X^i}_l,T^i_l)|0\rangle \nonumber\\
& &|out ;{\vec P^o}_1{\vec X^o}_1{ T^o}_1;-;{\vec P^o}_M{\vec X^o}_M{ T^o}_M\rangle \\
&=&A^{\dagger}({\vec P^o}_1,{\vec X^o}_1,T^o_1)A^{\dagger}({\vec P^o}_2,{\vec X^o}_2,T^o_2)--A^{\dagger}({\vec P^o}_l,{\vec X^o}_l,T^o_l)|0\rangle, 
\nonumber\\
& &T^i_l \ll T^o_l 
\end{eqnarray}
The differential transition probability from an initial state to a final
state is given as,
\begin{equation}
\label{eq:probability}
dP= \Pi_{l=1}^{l=L} {d{\vec P}_l d{\vec X}_l \over (2\pi )^3}|S_{out,in}|^2,
\end{equation}
and the total transition probability is obtained by integrating
positions and momenta as,   
\begin{equation}
\label{eq:total-probability}
P=\int{1 \over L!} \Pi_{l=1}^{l=L} {d{\vec P}_l d{\vec X}_l \over (2\pi )^3}|S_{out,in}|^2
\end{equation}
In the interaction picture, the S-matrix element is computed by
\begin{equation}
\label{eq:smatrix}
\langle 0|\Pi_{l=1}^{L} A({\vec P}_l,{\vec X}_l,T_l) T \exp 
\int dt'{ H_{int}(t') \over i}\Pi_{m=1}^{M} A^{\dagger}({\vec P}_m,{\vec X}_m,T_m)|0\rangle.
\end{equation}

As will be seen in examples of the next section, the differential probability
depends upon  the sizes of wave packets in the initial states and final
states. But total probabilities
become universal values that are independent from the sizes of wave
packets owing to the completeness of the many body states.

 Firstly, the
total probability from one initial state to a final state of fixed number of 
particles  become independent from the wave packet sizes of final
states. To see this, let define a generalized S-matrix of wave packet 
size $\sigma_o$  of the final state and  $\sigma_i$ of the initial state, 
\begin{eqnarray}
& &S_{out,in}(\sigma_o,\sigma_i)  \\
&= &\langle {\vec P^o}_1{\vec X^o}_1{ T^o}_1,\sigma_o;-;{\vec P^o}_L{\vec X^o}_L{ T^o}_L\sigma_o;out |in ;{\vec P^i}_1{\vec X^i}_1{ T^i}_1\sigma_i;-;{\vec P^i}_M{\vec X^i}_M{ T^i}_M\sigma_i\rangle. \nonumber
\end{eqnarray}

Using complexness relation of wave packets for an arbitrary wave packet size
, the total probability from one initial state to a L-particle state of
one value of $\sigma_o$ satisfy
\begin{eqnarray}
& &P_M(\sigma_0,\sigma_i) \\
& &=
{1 \over L!}\Pi_{l=0}^{l=L}\int {d{\vec P}_l d{\vec X}_l \over (2 \pi )^3}
\langle in; \sigma_i |S|L, {\vec P}_l,{\vec X}_l,out; \sigma_o\rangle  \langle L,{\vec P}_l,{\vec X}_l out; \sigma_o |S^{\dagger}|in;\sigma_i \rangle  \nonumber \\
& &={1 \over L!}\Pi_{l=0}^{l=L}\int {d{\vec P}_l d{\vec X}_l \over (2 \pi )^3}
\langle in; \sigma_i|S|L,{\vec P}_l,{\vec X}_l,out;\sigma_o'\rangle \langle L,
{\vec P}_l,{\vec X}_l,out;\sigma_o' |S^{\dagger}|in;\sigma_i \rangle \nonumber \\ 
& &=P_M(\sigma_0',\sigma_i).\nonumber 
\end{eqnarray}
Thus the total probability is independent from the size $\sigma_o$.
The above total probability also agrees with the total probability from
the initial state of  wave packets  to the final states of  momentum 
eigenstates,
\begin{eqnarray}
& &P_M(\sigma_0,\sigma_i) \\
& &={1 \over L!}\Pi_{l=0}^{l=L}\int {d{\vec p}_l \over (2 \pi )^{(3/2)}}
\langle in; \sigma_i |S|L, p_l, out;  \rangle  \langle L,p_l, out;  |S^{\dagger}|in;\sigma_i \rangle  \nonumber \\
& &=P_M( momentum state,\sigma_i).\nonumber
\end{eqnarray} 
In the above derivations, we have used the fact that the set of
N-particle states is complete regardless of the wave packet sizes.  Thus 
the scattering amplitude for any complete set of functions gives the
same total probability. Conversely we can compute the total probability
by using the momentum eigenstates for the unobserved particles as far as
the boundary conditions are satisfied.  

Secondly, the total probability from one state to any
possible final states,
\begin{eqnarray}
& &\sum_L P(in \rightarrow L) \\
\label{eq:phasespace}
& &=\sum_L {1 \over L!}\Pi_{l=0}^{l=L}\int {d{\vec P}_l d{\vec X}_l \over (2 \pi )^3}
\langle in |S|L,{\vec P}_l,{\vec X}_l,\sigma_o;out \rangle \langle L,{\vec P}_l,{\vec X}_l,\sigma_o;out   |S^{\dagger}|in\rangle \nonumber \\
& &=\langle in|SS^{\dagger}|  in \rangle \nonumber \\
& &=1 \nonumber
\end{eqnarray}    
becomes unity. The facts that the particle states are normalized and $S$
is unitary are used in the above derivation.  Thus the standard
probability interpretation for the square of the absolute value of the 
amplitudes is applicable with the phase space defined in 
Eq.(\ref{eq:phasespace}) despite  of the non-orthogonality of the 
states.

Finally an inclusive  probability where a partial set of
kinematical variables are measured and other variables are unmeasured 
satisfies  also a similar universal relation as 
the total probability. Namely this probability depends on the size of wave
packet of measured particles but does not depend upon the sizes of wave
packets of unmeasured particles, if the different values of wave packet
sizes are used. 
 
It is summarized as follows: The probability depends upon the
sizes of wave packets of measured particles and the probability does not 
depend on the variables of unmeasured particles, such as the momenta,
positions and wave packet sizes.    
 
\section{ Few body scattering }

Few  body scattering are studied as  examples. We study
a scattering process where a particle of a momentum ${\vec P}_1$ 
which is prepared at a space and time  coordinate ${\vec X}_1,T_1$ and
another particle of a momentum ${\vec P}_2$  which is prepared at 
${\vec X}_2,T_2$ collide and a particle of  a momentum 
${\vec P}_3$ at ${\vec X}_3,T_3 $ and another 
particle of a momentum ${\vec P}_4$ at ${\vec X}_4,T_4 $ are  measured first. 

A system with an interaction Hamiltonian,
\begin{equation}
\label{eq:interaction Hamiltonian}
H_{int}=\int d{\vec x} {\lambda \over 4}{\phi(x)}^4
\end{equation} 
is studied. This  interaction is of short range, and the generalized 
amplitude shows characteristic dependences on  the momentum as well as 
coordinate. 
\subsection{effective sizes of the interaction region }
Let substitute the
expansion of the 
field Eq.(\ref{eq:field expansionA}) and the interaction Hamiltonian
Eq.(\ref{eq:interaction Hamiltonian}) into Eq.(\ref{eq:smatrix}). Then we have
the scattering matrix in the first order of $H_{int}$,
\begin{eqnarray}
& &\langle 0|\Pi_{l=1}^{2} A({\vec P}_l,{\vec X}_l,T_l)  
\int dt'{ H_{int}(t') \over i}\Pi_{m=1}^{2} A^{\dagger}({\vec P}_m,{\vec X}_m,T_m)|0\rangle\nonumber  \\
& &=\lambda \int dt d{\vec x} \Pi_l C({\vec P}_l,{\vec X}_l,T_l,{\vec x},t) 
\Pi_m (C( {\vec P}_m,{\vec X}_m,T_m, {\vec x},t))^*.
\end{eqnarray} 
We have used the two point function of  the field of mixed 
representation and the field of coordinate representation, 
\begin{eqnarray}
& &\langle 0|\phi({\vec  x},t) A^{\dagger}({\vec P},{\vec X},T)|0\rangle \nonumber \\
& &=\int {d{\vec P}_0 d{\vec X}_0 \over (2\pi )^3}C({\vec P}_0,{\vec X}_0,
{\vec x},t)  \langle 0| A({\vec P}_0,{\vec X}_0,t) A^{\dagger}({\vec P},{\vec X},T)|0\rangle.
\end{eqnarray}
This is written further by combining Eq.(\ref{eq:field expansionA}) and Eq.(\ref{eq:Ctransformation}) as, 
\begin{equation}
\langle 0|\phi({\vec  x},t) A^{\dagger}({\vec P},{\vec X},T)|0\rangle=C({\vec P},{\vec X},T,{\vec x},t).  
\end{equation}
In the parameter  regions we are interested in  this paper, this
function and its partner is approximated well with a very  good
accuracy as,
\begin{eqnarray}
& & C({\vec P},{\vec X},T,{\vec x},t)  ={1 \over \sqrt{2E({\vec P})}}\langle {\vec P},{\vec X},T|{\vec x},t \rangle \\
& & \tilde C({\vec P},{\vec X},T,{\vec x},t)  = \sqrt{2E({\vec P})}\langle {\vec P},{\vec X},T|{\vec x},t \rangle 
\end{eqnarray}
In the following calculations we use these formula.


The coefficients $C({\vec P}_l,{\vec X}_l,T_l,{\vec x},t) $ and their
complex conjugate  give the
values of wave functions at $({\vec x},t)$. In the region where times
$T_l$ are close to the $t$, the product of the functions is a Gaussian 
function around the peak in the variables ${\vec x}$ and $t$ and is
expressed as 
\begin{eqnarray}  
& &\Pi_l C({\vec P}_l,{\vec X}_l,T_l,{\vec x},t) 
\Pi_m (C( {\vec P}_m,{\vec X}_m,T_m, {\vec x},t))^*\\
&=&(N_3)^L \Pi \exp{(-{1 \over 2\sigma}({\vec x}-{\vec X}_j-{\vec v}_j(t-T_j))^2)} \times \nonumber\\
& &
\exp{(-iE({\vec P}_l)(t-T_l)+i{\vec P}_l\cdot({\vec x-{\vec X}_l}) 
+iE({\vec P}_m)(t-T_m)-i{\vec P}_m\cdot({\vec x-{\vec X}_m}))} \nonumber\\
&=&N \exp({-{1 \over 2 \sigma_{S}}({\vec x}-{\vec x}_0)^2-{1 \over 2 \sigma_{T}}(t-t_0)^2})e^{i\phi}.\nonumber 
\end{eqnarray}
 The peak position ${\vec x}_0$ and $t_0$ are determined as,
\begin{eqnarray}
& &{\vec x}_0= \sigma_S \sum {1 \over \sigma_l}{\vec x}_l(t),           \\
& &t_0={B \over \sigma_T},                    \\
& &B=\sum_l{1 \over \sigma_l}({\vec X}_l-{\vec v}_l T_l)\cdot{\vec v}_l-\sigma_S(\sum_l{1 \over \sigma_l}{\vec v}_l)\cdot(\sum_l{1 \over \sigma_l}({\vec X}_l-{\vec v}_lT)),   \\
& &{\vec x}_l(t)={\vec X}_l+{\vec v}_l(t-T_l) 
\end{eqnarray}
and the variances $\sigma_{S}$ and $\sigma_{T}$ are determined as 
\begin{eqnarray}
& & \sigma_S=  {\sigma \over 4}               \\
& &\sigma_T=   \sigma \left(\sum_l{{\vec v}_l}^2-{1 \over 4}(\sum_l{\vec v}_l)^2 \right)^{-1}.                 
\end{eqnarray}
The  region around the peak within the 
spatial distance $\sqrt \sigma_S$ and the time distance $\sqrt \sigma_T$ 
gives a dominant contribution to the integral. 
Both distances are proportional to the $\sigma$. So if the $\sigma_S $ is
small, the $\sigma_T$ is also small. The normalization $N$
and the phase $\phi$ are complicated functions of the energies, momenta,
spatial positions, and temporal positions. Explicit formulas are given
in Appendix.
\subsubsection{Complete measurements}
So far, all particles have the same wave packet size and all particles
are measured. When this is not hold and different particles have
different  wave packet sizes and some particles are not measured, the 
effective sizes $\sigma_S $ and  $\sigma_T$ become different from the
above values. We study
the behavior of these variances in  general cases where each wave
packet has its own size and all momenta and positions are
measured here. Let specify the wave packet size of the   
l-th particle as $\sigma_l$, and its velocity as ${\vec v}_l$ , then the 
variances $\sigma_{S}$ and $\sigma_{T}$ are given by
\begin{eqnarray}
& & \sigma_S=\left( \sum_l{1 \over \sigma_l}\right)^{-1}                 \\
& &\sigma_T=\left(\sum_l{1 \over \sigma_l}{{\vec v}_l}^2-\sigma_S(\sum_l{1 \over \sigma_l}{\vec v}_l)^2 \right)^{-1}                   
\end{eqnarray}
The $\sigma_S$ is determined mainly by the small $\sigma_l$ of the
measured particles but the
$\sigma_T$ is determined by the $\sigma_l$ and ${\vec v}_l$ of the
measured particles. The
small $\sigma_l$ does not contribute if the corresponding ${\vec v}_l$ 
vanishes. So the time variance $\sigma_T$ could become
large even though the space variance $\sigma_S$ is small. The large
$\sigma_t$ is compatible with the small $\sigma_S$ in this case. 

 The ${\vec v}_l$ depends on the momentum ${\vec P}_l$. So $\sigma_T$
 is not a constant generally but varies in the kinematical region of the
 final state. The exception is the case when $\sigma_l$ is infinity,
i.e., plane wave.   

\subsubsection{Partial measurements}
When some portion of particles are measured and others are unmeasured,
the probability depends on the wave packet sizes of the measured
particles. 
When one particle, $l=1$, is measured and other particles are 
unmeasured,it depends on the wave packet $ \sigma_1$. Hence the
effective sizes of the vetcex area for computing the total probability
are obtained by letting   $\sigma_l=\infty$ for $l \ne 1$
\begin{eqnarray}
& & \sigma_S= \sigma_1                \\
& &\sigma_T=({1 \over \sigma_1}{{\vec v}_1}^2-{1 \over \sigma_1}{\vec v}_1^2 )^{-1}   =\infty                 
\end{eqnarray}
The spatial size $\sigma_S$ is determined from the $\sigma_1 $ of the
observed particle but the temporal size $\sigma_T$ diverges. 
    
Next, the probability when  two particles are measured and other particles are 
unmeasured depends on the wave packet $ \sigma_1 $ and $ \sigma_2 $ and
velocities $ {\vec v}_1$ and $ {\vec v}_2 $
of the measured particles. 
The effective sizes are obtained by letting   $
\sigma_l=\infty$ for $l \ne 1,2$

\begin{eqnarray}
& & \sigma_S=\left( \sum_{l=1}^{l=2}{1 \over \sigma_l}\right)^{-1}              ={ \sigma_1\sigma_2 \over \sigma_1+ \sigma_2}   \\
& &\sigma_T=\left(\sum_{l=1}^{l=2}{1 \over \sigma_l}{{\vec v}_l}^2-\sigma_S(\sum_{l=1}^{l=2}{1 \over \sigma_l}{\vec v}_l)^2 \right)^{-1}={  \sigma_1+ \sigma_2 \over ({\vec v}_1-{\vec v}_2)^2 }                    
\end{eqnarray}
    
$\sigma_T$ diverges if $ {\vec v}_1$ is equal to ${\vec v}_2$.

\subsection{Short distance scattering }
When two particles are in the initial states and two particles are in the final
states and the distances $|{\vec X}_i-{\vec X}_j|$ and $ |T_i-T_j| $ are 
small, the formula Eq.($\ref{eq:closetime}$) for the small time
differences is used. We expect that the amplitude in this region shows 
features of the translational motion of wave packets and other features of the
generalized scattering amplitude. For simplicity, we present the results
when  all the particles have the same wave packet size.  The general
case is given in the Appendix.   

The transition amplitude in the lowest order of $H_{int}$ is given by 
\begin{eqnarray}
& &\lambda \int dt d{\vec x} \Pi_l C({\vec P}_l,{\vec X}_l,T_l,{\vec x},t) 
 \Pi_m (C(  {\vec x},t,{\vec P}_m,{\vec X}_m,T_m ,{\vec x},t))^*\\
& &=\lambda N_3^4 \Pi_l ({2 E({\vec P}_l)})^{-1/2}\int dt d{\vec x} 
\exp\sum \{-is_i\{(t-T_l)E({\vec P}_l)-{\vec P}_l\cdot({\vec x}-
{\vec X}_l)\}\nonumber \\
& &\exp\sum\{ -{1 \over 2\sigma}
\{{\vec x}-{\vec X}_l-{\vec v}_l(t-T_l) \}^2 \}, \nonumber \\
& &=\tilde N \exp(i\phi -R), \nonumber  \\
& &\tilde N=({4 \over \sum_{i,j}({\vec v}_i-{\vec v}_j)^2})^{3/2} \Pi_l ({2 E({\vec P}_l)})^{-1/2},      \\
& &\phi=\sum_i sgn_i(T_iE({\vec P}_i)-{\vec P}_i \cdot{\vec X}_i)+
\tilde \phi,  \\
& &R={\sigma \over 8}(\sum_i {\vec P}_i)^2+{1 \over 8\sigma}
\sum_{i,j}({\vec X}_i-{\vec v}_iT_i- {\vec X}_j+{\vec v}_jT_j)^2 \nonumber\\
& &+{2 \sigma \over \sum_{i,j}({\vec v}_i-{\vec v}_j)^2}
(\sum E({\vec P}_i)s_i )^2 +\tilde R,
\end{eqnarray}
where $sgn_i$ is a signature  
\begin{eqnarray}
sgn_i= +1(in-state),-1(out-state)
\end{eqnarray}  
and $\tilde \phi$ and $\tilde R$ are small
quantities and are given in the appendix for most general case.
In the above equations, the dominant part in the phase,  $\phi$, is the 
standard phase of the plane wave of the momentum ${\vec p}_0$ and energy 
$E_0$. From the first factor and
third factor of normalization
$R$, the momentum conservation is 
approximately satisfied with the variance, $({\sigma \over 8})^{-1/2}$
and the energy conservation is also approximately satisfied with the
variance  $({2 \sigma \over \sum_{i,j}({\vec v}_i-{\vec v}_j)^2})^{-1/2}$.
The second factor of the $R$ shows that the particle trajectories
coincide and coordinates ${\vec X}_i-{\vec v}_iT_i$ are the same within
the distance ${(8\sigma)}^{1/2}$. So particles follow classical
trajectories.

From the amplitudes, we define new transition probabilities that depend
upon the coordinates in addition to the momenta and argue on an
asymptotic condition of the standard scattering amplitude which depends
upon the momenta. 

The transition probability is a square of the absolute value of the
amplitude and is expressed as 
\begin{eqnarray} 
& &P({\vec P}_3,{\vec X}_3,T_3;{\vec P}_4,{\vec X}_4,T_4) \\
& &={1 \over 2!} {1 \over (2\pi )^3}(\tilde N)^2 \exp(-{\sigma \over 4}(\sum_i {\vec P}_i)^2) 
\nonumber \\
& &\exp(-{1 \over 4\sigma}\sum_{i,j}({\vec X}_i-{\vec v}_iT_i- {\vec X}_j+{\vec v}_jT_j)^2 -{4 \sigma \over \sum_{i,j}({\vec v}_i-{\vec v}_j)^2}(\sum E({\vec P}_i)s_i )^2  ). \nonumber 
\end{eqnarray}
This has a peak at the positions where the
conditions
\begin{equation}
{\vec X}_i-{\vec v}_iT_i- {\vec X}_j+{\vec v}_jT_j=0
\end{equation}
are satisfied. Thus the  peak is along a line  
\begin{equation}
{\vec X}_i={\vec v}_i T_i+{\vec C},
\end{equation}
where ${\vec C}$ is a constant vector. These positions depend upon the
times and the time $T_i$
are arbitrary in the present formalism, hence it is possible to choose
the times in such manner that these positions are inside of detectors if
the detector is located in
the direction of the momentum. Then total probability integrated on this
direction is measured. To see this probability, let us decompose the 
position vector of the i-th 
particle into the longitudinal component and the transverse components
with respects to the velocity ${\vec v}^i$,
\begin{eqnarray}
& &{\vec X}^i={\vec v}^i S^i +{{\vec n}^i}_T {X^i}_T,\\ 
& &{\vec v}^i \cdot {{\vec n}^i}_T=0,   \\
& &   {{\vec n}^i}_T \cdot {{\vec n}^i}_T=\delta_{ij}.
\end{eqnarray}
The volume element is written as 
\begin{equation}
d{\vec X}^j=d S^jd {X^i}_T | v^i|. 
\end{equation}

Using these variables, the Gaussian factor of the differential
probability is written as 
\begin{eqnarray}
& &\sum_{ij}({\vec X}^i-{\vec v}^i T^i-{\vec X}^j+{\vec v}^j T^j)^2 \\
& &=\sum_{ij}({\vec v}^i(S^i-T^i)+{\vec n}^i_TX^i_T -{\vec v}^j(S^j-T^j)-{\vec n}^j_TX^j_T)^2 \nonumber\\
& &=\sum_{ij}({\vec v}^i \tilde S^i+{\vec n}^i_TX^i_T -{\vec v}^j \tilde S^j-{\vec n}^j_TX^j_T)^2, \nonumber  \\
& &\tilde S^i= S^i-T^i. 
\end{eqnarray}
Thus the longitudinal variable $S_i$ is combined with time $T^i$ and
it is possible to replace  the longitudinal
coordinate with the time variable in Eq.(\ref{eq:probability}). 

 The transition probability is given by  an integration of a differential 
probability over the momenta and coordinates  as,
\begin{equation}
\label{eq:probability}
P=\int \Pi_{m=3}^{m=4} {d{\vec P}_m d{\vec X}_m  \over (2 \pi )^3}
P({\vec P}_3,{\vec X}_3,T_3;{\vec P}_4,{\vec X}_4,T_4).
\end{equation}

In the ordinary detectors  neither the the precise value of the time
$T^i$ nor the longitudinal coordinates ${X^i}_L$
are  measurable but the total probability $P({\vec P}^3,{\vec X}^3_T,;{\vec
P}^4,{\vec X}^4_T) $ integrated on these variables is measured.

To obtain this probability,the variable $S_i$ is integrated. Then  the 
probability $P({\vec P}^3,{\vec X}^3_T,T^3;{\vec P}^4,{\vec X}^4_T,T^4)$
is found and is written as 
\begin{eqnarray}
& & P({\vec P}^3,{\vec X}^3_T,T^3;{\vec P}^4,{\vec X}^4_T,T^4) \\
& &= \int \Pi_{m=3}^{m=4}  d{\vec X}^m_L {1 \over (2 \pi )^{1/2}}
P({\vec P}^3,{\vec X}^3,T^3;{\vec P}^4,{\vec X}^4,T^4) \nonumber \\
& &=P({\vec P}^3,{\vec X}^3_T,;{\vec P}^4,{\vec X}^4_T).\nonumber 
\end{eqnarray}
The time dependence disappears in $P({\vec P}^3,{\vec X}^3_T,T^3;{\vec P}^4,{\vec X}^4_T,T^4)$.

Next, we make a connection of the present result with a standard
scattering matrix where an asymptotic condition is satisfied and only
the momenta are observed.    
In the ordinary scattering processes the initial time $T_i$ is 
$-\infty$ and the final time is $+\infty$. The distance $|{\vec
X}_i-{\vec X}_j|$ in the initial state is proportional to 
$|{\vec v}_i-{\vec v}_j|T_i$ and the distance $|{\vec X}_i-{\vec X}_j|$
in the final state is proportional to $|{\vec v}_i-{\vec v}_j|T_i$. They 
become large except $|{\vec v}_i-{\vec v}_j|=0$. When we define this
case from a limit
$|{\vec v}_i-{\vec v}_j| \rightarrow 0$ where a large $T$
limit is taken first, a distance between two wave packets becomes
infinity and the wave packets at $T \rightarrow \pm \infty$ do not
overlap each others.  The theory thus defined satisfies asymptotic
condition.
   
By integrating  the coordinates, we have the
momentum dependent differential probability,
\begin{eqnarray}
\label{eq:mprobability}
& &P({\vec P}_3,;{\vec P}_4)=N \exp({-{\sigma \over 4}({\sum}_i {\vec P}_i)^2 -{4 \sigma \over \sum_{i,j}({\vec v}_i-{\vec v}_j)^2 }(\sum E_is_i)^2 })  \\
& &N={1 \over 2} (4\sigma \pi)^32^{-3/2}\exp({-{3 \over 8\sigma}({\vec X}_1-
{\vec v}_1 T_1-{\vec X}_2+{\vec v}_2 T_2)^2}).
\end{eqnarray}
In Eq.({\ref{eq:mprobability}}), the normalization factor $N$ is a
constant which  does not depend on the final state and the  momentum dependent
probability is almost the same as the probability of the plane
waves. By integrating momenta of the final states,we have  the total 
probability.

If the initial state and the final state have different values of the
$\sigma$, we use the formula given in the appendix.
Let $\sigma_o$ be the size for all final particles and the $\sigma_i$
be the size for all initial particles, the probability is given
as,
 \begin{eqnarray}
\label{eq:mprobability2}
& &P({\vec P}_3,;{\vec P}_4)=N \exp({-\sigma}_s ({\sum}_i {\vec P}_i)^2 -
\sigma_t (\sum E_is_i)^2 )  \\
& &N={1 \over 2} (4 \pi)^3({\sigma_i \sigma_o \over 2})^{3/2}  \exp({-{3 \over 8\sigma_i}({\vec X}_1-
{\vec v}_1 T_1-{\vec X}_2+{\vec v}_2 T_2)^2}), \\
& &\sigma_s={1 \over 2}({1 \over \sigma_o}+{1 \over \sigma_i} )^{-1}     ,\\
& &\sigma_t=(\Sigma_j{ {\vec v_j}^2 \over \sigma_j}-{{\vec v}_0^2 \over \sigma_s}
 )^{-1} .
\end{eqnarray} 

\subsection{Long distance scatterings:first order }
When one of the times,  ${T_1}$, is in a position far away from other 
times ${T_l(l \ne 1)}$, and the classical trajectories  meet at around a 
time near $T_l(l \ne 1)$ and coordinate ${\vec X}_l(l \ne 1) $, the
dominant contribution in the integration comes from the region  
near ${\vec X}_l(l \ne 1)$ and the position ${{\vec X}_l-{\vec
v}_lT_l}$. One of the time difference $t-T_l$ becomes large and
asymptotic expansion for the corresponding $C({\vec P}_1,{\vec
X}_1,T_1|t,{\vec x})$ is used. The transition matrix element becomes,then,
\begin{eqnarray}
\label{eq:remote-time}
& &\langle 0|\Pi_{l=1}^{2} A({\vec P}_l,{\vec X}_l,T_l)  
\int dt'{ H_{int}(t') \over i}\Pi_{m=1}^{2} A^{\dagger}
({\vec P}_m,{\vec X}_m,T_m)|0\rangle \nonumber\\
& &=\lambda \int dt d{\vec x} \Pi_lC( {\vec P}_l,{\vec X}_l,T_l|{\vec x},t) 
\Pi_mC({\vec P}_m,{\vec X}_m,T_m|{\vec x},t)^{*} \nonumber \\
& &=\lambda ({1 \over 2E({\vec P}_1)2E({\vec P}_2)2E({\vec P}_3)2E({\vec P}_4) } )^{1/2 } \int dt d{\vec x}N_{asym}^{*}e^{iE({\vec P}_{({\vec X}_1-{\vec x})})(t-T_1)-i{\vec P}_1({\vec x}-{\vec X}_1)} \nonumber \\
& &{N_{3}}^{*}e^{-{1 \over 2 \sigma}({\vec x}-{\vec X}_2-{\vec v}_2(t-T_2))^2 +iE({\vec P}_2)(t-T_2)-i{\vec P}_2({\vec x}-{\vec X}_2)}\\
& & \times \Pi_{j=3,4} {N_{3}}e^{-{1 \over 2 \sigma}({\vec x}-{\vec X}_j-{\vec v}_j(t-T_j))^2 -iE({\vec P}_j)(t-T_j)+{\vec P}_j({\vec x}-{\vec X}_j)} \nonumber 
\end{eqnarray} 
where $N_{asym}$ and the stationary momentum ${\vec P}_{{\vec
X}_1-{\vec x}}$ are 
\begin{eqnarray}
\label{eq:asymptotic-s}
& &N_{asym}=N_3({1 \over {2i {\gamma}_L \over \sigma}+1})^{1/2}({1 \over {2i {\gamma}_T \over \sigma}+1})e^{-{1 \over 2}\sigma ({\vec P}_X -{\vec P}_1)^2}\\
& &{\vec P}_{{\vec X}_1-{\vec x}}=({\vec X}_1-{\vec x}) {m \over ((T_1-t)^2-({\vec X}_1-{\vec x})^2)^{1/2}}   \nonumber
\end{eqnarray} 
Substituting these expressions, we have 
\begin{eqnarray}
\label{eq:asymptotic1}
& & \lambda \int dt d{\vec x} \Pi_lC( {\vec P}_l,{\vec X}_l,T_l|{\vec x},t) 
\Pi_mC({\vec P}_m,{\vec X}_m,T_m|{\vec x},t)^{*}  \nonumber \\
& &=\lambda ({1 \over 2E({\vec P}_1)2E({\vec P}_2)2E({\vec P}_3)2E({\vec P}_4) } )^{1/2 } (|N_3|^2)^2 ({1 \over {2i {\gamma}_L \over \sigma}+1})^{1/2}
({1 \over {2i {\gamma}_T \over \sigma}+1}) \nonumber\\
& &\int dt d{\vec x}
e^{im \sqrt{(t-T_1)^2-({\vec x}-{\vec X}_1)^2}}  \exp({ -{1 \over 2}\sigma  {(E({\vec P}_0))^2 \over (t-T_1)^2}{({\vec x}-{\vec x}_0 )_T}^2 -{1 \over 2}\sigma {(E({\vec P}_1))^6 \over m^4 (t-T_1)^2}{({\vec x}-{\vec x}_0)_L}^2}) 
\nonumber \\
& &\tilde N e^{i\tilde \phi}\exp{(-{1 \over 2 \sigma_s}({\vec x}-{\vec x}_0')^2-{1 \over 2 \sigma_t}({t}-{t}_0')^2 )},
\end{eqnarray}
where the center position ${\vec x}_0$ is given by
\begin{equation}
{\vec x}_0={\vec X}_1+(t-T_1){{\vec P}_1 \over E({\vec P}_1)},
\end{equation}
and the normalization , the phase, the  variances, and center positions
are defined by diagonalizing the  products of wave packets,
\begin{eqnarray}
& &\tilde N e^{\tilde \phi }\exp{(-{1 \over 2 \sigma_s}({\vec x}-{\vec x}_0')^2-{1 \over 2 \sigma_t}({t}-{t}_0')^2 )}={N_{3}}^{*}e^{-{1 \over 2 \sigma}({\vec x}-{\vec X}_2-{\vec v}_2(t-T_2))^2 +iE({\vec P}_2)(t-T_2)-i{\vec P}_2({\vec x}-{\vec X}_2)}\nonumber \\
& & \times \Pi_{j=3,4} {N_{3}}e^{-{1 \over 2 \sigma}({\vec x}-{\vec X}_j-{\vec v}_j(t-T_j))^2 -iE({\vec P}_j)(t-T_j)+{\vec P}_j({\vec x}-{\vec X}_j)}.  
\end{eqnarray} 
If the variances $\sigma_s$ and $\sigma_t$ are small values, the
amplitude is further written as,
\begin{eqnarray}
& &\lambda ({1 \over 2E({\vec P}_1)2E({\vec P}_2)2E({\vec P}_3)2E({\vec P}_4) } )^{1/2 } (|N_3|^2)^2 ({1 \over {2i {\gamma}_L \over \sigma}+1})^{1/2}
({1 \over {2i {\gamma}_T \over \sigma}+1}) \nonumber\\
& &
e^{im \sqrt{(t_0'-T_1)^2-({\vec x}_0'-{\vec X}_1)^2}}  \exp({ -{1 \over 2}\sigma  {(E({\vec P}_0))^2 \over (t_0'-T_1)^2}{({\vec x}_0'-{\vec x}_0 )_T}^2 -{1 \over 2}\sigma {(E({\vec P}_1))^6 \over m^4 (t_0'-T_1)^2}{({\vec x}_0'-{\vec x}_0)_L}^2}) 
\nonumber \\
& &\tilde N e^{i\tilde \phi}  (2\sigma_s \pi)^{3 \over 2} (2 \sigma_t)^{1 
\over 2}.
\end{eqnarray}
This expression of the amplitude shows that the wave packet expands and
has the phase factor which is proportional to the square root of the
proper time.
  
When we integrate on the variables ${\vec x},t$ first in Eq.$(\ref{eq:remote-time})$,
we have, 
\begin{eqnarray}
\label{eq:asymptotic2}
& &\lambda \int dt d{\vec x} \Pi_lC( {\vec P}_l,{\vec X}_l,T_l|{\vec x},t) 
\Pi_mC({\vec P}_m,{\vec X}_m,T_m|{\vec x},t)^{*}  \\
& &=(|N_3|^2)^2 ({2\sigma \pi \over 3})^{4/2}({1 \over <({\vec v})^2>-<{\vec v}>^2 })^{1/2} \tilde N \nonumber \\
& &\int d{\vec p} e^{-iE({\vec p}) (T_1-\delta T_1)+i{\vec p}({\vec X}_1+\delta {\vec X}_1)-{\sigma' \over 2}({\vec p}-{\vec P}_1-\delta{\vec P}_1)^2}, \nonumber \\
& &<{\vec v}>={1 \over 3}\sum_{j=1,3} {\vec v}_j, \\
& &<{\vec v}^2>={1 \over 3}\sum_{j=1,3} ({\vec v}_j)^2, 
\end{eqnarray}
where $\delta T_1$, $\delta {X_1}$, and ${\delta {\vec P}_1}$ are of the 
order $O(1)$. $\tilde N$ is a normalization factor which depends upon
kinematical variables. The integration on the variable ${\vec p}$ is
carried with a use of  the stationary phase approximation as in the
previous cases.


Finally we have the amplitude,
\begin{eqnarray}
\label{eq:asymptotic3}
& &\lambda \int dt d{\vec x} \Pi_lC( {\vec P}_l,{\vec X}_l,T_l|{\vec x},t) 
\Pi_mC({\vec P}_m,{\vec X}_m,T_m|{\vec x},t)^{*}  \nonumber \\
& &=(|N_3|^2)^2 ({2\sigma \pi \over 3})^{4/2}({1 \over <({\vec v})^2>-<{\vec v}>^2 })^{1/2} \tilde N \\
& &e^{-im\sqrt{(T_1-\delta T_1)^2-{({\vec X}_1+\delta {\vec X}_1)^2 }}
 -{1 \over 2}\sigma  {(E({\vec P}_1))^2 \over (T_1-\delta T_1)^2}{({\vec X}_1-{\delta \vec X}_1 )_T}^2 -{1 \over 2}\sigma {(E({\vec P}_1))^6 \over m^4 (T_1-\delta T_1)^2}{({\vec X}_1-\delta {\vec X}_1)_L}^2}.\nonumber
\end{eqnarray}
Thus the amplitude depends on the large variables ${\vec
X}_1$ and $T_1$ in a simple  form. The normalization factor is
inversely proportional to $T_1$ and the phase 
factor is proportional to the mass and the proper time 
$m \sqrt {c^2(T_1)^2-({\vec X}_1)^2}$.    

\subsection{Long distance scattering :second order }
Next we study the few  body scattering amplitude in the second order of interaction
where there is one propagator $D(t,{\vec x})$. The propagator  connects two
interaction points,$({\vec x}_1,t_1)$ and $({\vec x}_2,t_2)$,  
\begin{eqnarray}
& &\langle 0|\Pi_{l=1}^{3} A({\vec P}_l,{\vec X}_l,T_l)  
\int dt_1 dt_2 {T( H_{int}(t_1)H_{int}(t_2)) \over i^2 } \Pi_{m=1}^{3} A^{\dagger}
({\vec P}_m,{\vec X}_m,T_m)|0\rangle\nonumber \\
& &={\lambda}^2 \int dt_1 d{\vec x}_1  \int dt_2 d{\vec x}_2
V_g(t_1,{\vec x}_1,{\vec P}_l,\cdots)
D(t_1-t_2,{\vec x}_1-{\vec x}_2)
V_g(t_2,{\vec x}_2,{\vec P}_2,\cdots)^{*}, \nonumber\\ 
& &V_g(t_1,{\vec x}_1,{\vec P}_l,\cdots)=\Pi_lC( {\vec P}_l,{\vec X}_l,T_l|{\vec x}_1,t_1), \\
& &V_g(t_2,{\vec x}_2,{\vec P}_2,\cdots)^{*}=\Pi_mC({\vec P}_m,{\vec X}_m,T_m|{\vec x}_2,t_2)^{*}.\nonumber
\end{eqnarray}
The propagator   $D(t_1-t_2,{\vec x}_1-{\vec x}_2)$ is given by,
\begin{equation}
D(t_1-t_2,{\vec x}_1-{\vec x}_2)=i \int {d^3 p \over (2\pi)^3 2E({\vec p})} e^{ip(x_1-x_2)}|_{E({\vec p})=\sqrt{{\vec p}^2+m^2}}.
\end{equation}  
We study the configuration when  times $T_l(l=1,3)$ are close each others
and times $T_m(m=4,6)$ are close each others but the first group of
times is  separated from the second group of times with a large
distance. Regions when the time variable $t_1$ is near $T_l(l=1,3)$ and
the other time variable $t_2$ is near $T_m(m=4,6)$ or the opposite give the 
dominant contribution to the amplitude in the time integration. Since
the distance $|t_1-t_2|$ is large,  on mass shell kinematical region where 
$p^2=m^2$ is satisfied is dominant in the momentum integration. 
Using the stationary phase approximation in the momentum integration, 
we replace the propagator with the asymptotic form that is obtained at
the stationary momentum ${\vec p}_x$  
\begin{eqnarray}
& &D_{asym}(t_1-t_2,{\vec x}_1-{\vec x}_2)=i N_{x}{1 \over (2\pi)^3 2E({\vec p}_{x})} e^{ip_{x}(x_1-x_2)}|_{E({\vec p}_{x})}\\
& &N_{x}=({1 \over i\gamma_L})^{1/2}({1 \over i\gamma_T}),  \\
& &\gamma_T={1 \over 2}{(t_1-t_2) \over E({\vec p}_x)},\gamma_L=\gamma_T{m^2 \over E({\vec p}_x)^2 }, 
\end{eqnarray}
where the momentum ${\vec p}_x$ is given as,
\begin{equation}
\label{eq:stationary-momentum} 
{\vec p}_{\vec x}=({\vec x}_1-{\vec x}_2) 
{m \over ((t_1-t_2)^2-({\vec x}_1-{\vec x}_2)^2)^{1/2}}. 
\end{equation}
We substitute these expressions into the amplitude and we have,   
\begin{eqnarray}
& &{\lambda}^2 \int dt_1 d{\vec x}_1  dt_2 d{\vec x}_2  V_g(t_1,{\vec x}_1,{\vec P}_l,)D(t_1-t_2,{\vec x}_1-{\vec x}_2)V_g(t_2,{\vec x}_2,{\vec P}_m,{\vec X}_m,T_m)^{*} \nonumber \\
& &= {\lambda}   \int d {t_1} d{\vec x}_1 V_g(t_1,{\vec x}_1,{\vec P}_l,{\vec X}_l,T_l)  e^{ {ip_{x}{x_1}}}|_{E({\vec p}_{x})} iN_x({1 \over (2\pi)^3 2E({\vec p}_{x})})^{1/2}  \\
& & {\lambda} \int d {t_2} d {\vec x}_2 V_g(t_2,{\vec x}_2,{\vec P}_m,{\vec X}_m,T_m)^{*} e^{{-ip_{x} {x}_2}}|_{E({\vec p}_{x})} ({1 \over (2\pi)^3 2E({\vec p}_{x})})^{1/2}.  \nonumber                       
\end{eqnarray} 
In integrating $(t_i, {\vec x}_i), i=1,2$  the momentum ${\vec p}_{x}$ of 
Eq.(\ref{eq:stationary-momentum})
becomes a constant vector, if these variables are in the narrow
regions. Then the total
amplitude is   proportional to the product of the two amplitudes and is 
inversely proportional to the spreading of the wave packet $\gamma_T$.
The  total probability is proportional to the product of probabilities
of two processes and is inversely proportional to
${\gamma_T}^2$. Interference effect of propagating wave is negligible in 
this case.      
Thus in the present regime the intermediate state is treated 
as an observed particle. Hence the single particle treatment of the
intermediate state is applicable. Using the argument of
Ref.\cite{Stodolsky} and \cite{Kiers},
the effect of the wave packets are described by the ensemble of the
energy eigenstates in this regime. 

 In an opposite situation where these integration variables  cover wide regions
 and   ${\vec p}_{x}$ is not a constant vector but varies with these variables 
$(t_1,{\vec x}_1)$ or $(t_2,{\vec x}_2)$,  the
 naive single particle treatment is not justified. The total amplitude
 becomes linear combinations of the amplitudes of the various values of
 the momentum, ${\vec p}_{x}$, and become different from the product of
 two amplitudes. We split the integration regions into small regions $V_l$
 and obtain the momentum ${\vec p}_{x}^{ll'}$ defined from a pair of
 these regions. Using
 them we  have the amplitude,
\begin{eqnarray}
& &{\lambda}^2 \int dt_1 d{\vec x}_1  dt_2 d{\vec x}_2  V_g(t_1,{\vec x}_1,{\vec P}_l,)D(t_1-t_2,{\vec x}_1-{\vec x}_2)V_g(t_2,{\vec x}_2,{\vec P}_m,{\vec X}_m,T_m)^{*} \nonumber \\
& &= \sum_{l l'} {\lambda}   \int_{V_l} d {t_1} d{\vec x}_1 V_g(t_1,{\vec x}_1,{\vec P}_l,{\vec X}_l,T_l)  e^{ {ip_{x}^{ll'}{x_1}}}|_{E({\vec p}_{x}^{ll'})} iN_x({1 \over (2\pi)^3 2E({\vec p}_{x}^{ll'})})^{1/2}  \\
& & {\lambda} \int_{V_{l'}} d {t_2} d {\vec x}_2 V_g(t_2,{\vec x}_2,{\vec P}_m,{\vec X}_m,T_m)^{*} e^{{-ip_{x}^{ll'} {x}_2}}|_{E({\vec p}_{x}^{ll'})} ({1 \over (2\pi)^3 2E({\vec p}_{x}^{ll'})})^{1/2}.  \nonumber                       
\end{eqnarray} 
The total amplitude is a linear combination of the amplitudes of
different momentum ${\vec p}_{x}^{ll'}$. Probability may show the 
interference of the different intermediate momentum. 

 In this situation we are able to write the 
amplitude in a different manner. Let  write the propagator as,
\begin{equation}
D(t_1-t_2,{\vec x}_1-{\vec x}_2)=-2i \int d^3x D(t_1-t,{\vec x}_1-{\vec x}) \dot 
D(t-t_2,{\vec x}-{\vec x}_2).
\end{equation} 
We substitute this  expression and we have the amplitude,    
\begin{eqnarray}
& &{\lambda}^2\int dt_1 d{\vec x}_1  dt_2 d{\vec x}_2  V_g(t_1,{\vec x}_1,{\vec P}_l,)D(t_1-t_2,{\vec x}_1-{\vec x}_2)V_g(t_2,{\vec x}_2,{\vec P}_m,{\vec X}_m,T_m)^{*} \nonumber \\
& &= -2i{\lambda}^2 \int d{\vec x}  \int d {t_1} d{\vec x}_1 V_g(t_1,{\vec x}_1,{\vec P}_l,{\vec X}_l,T_l) D(t_1-t,{\vec x}_1-{\vec x}) \\
& &\times \int d{t_2} d{\vec x}_2 {\dot D(t-t_2,{\vec x}-{\vec x}_2)} V_g(t_2,{\vec x}_2,{\vec P}_m,{\vec X}_m,T_m)^{*}.  \nonumber                           \end{eqnarray} 

Using the stationary phase approximation in the momentum integration, 
we have the propagators, 
\begin{eqnarray}
& &D_{asym}(t_1-t,{\vec x}_1-{\vec x})=i N_{x}^{(1)}{1 \over (2\pi)^3 2E({\vec p}_{x}^{(1)})} e^{ip_{x}^{(1)}(x_1-x)}|_{E({\vec p}_{x}^{(1)})}\\
& &N_{x}^{(1)}=({1 \over i\gamma_L^{(1)}})^{1/2}({1 \over i\gamma_T^{(1)}}),  \\
& &\gamma_T^{(1)}={1 \over 2}{(t_1-t) \over E({\vec p}_x)},\gamma_L^{(1)}=\gamma_T^{(1)}{m^2 \over E({\vec P}_x^{(1)})^2 } \\
& &{\vec P}_{\vec x}^{(1)}=({\vec x}_1-{\vec x}) {m \over ((t_1-t)^2-({\vec x}_1-{\vec x})^2)^{1/2}}   \
\end{eqnarray}
and 
\begin{eqnarray}
& &\dot D_{asym}(t-t_2,{\vec x}-{\vec x}_2)=i^2 N_{x}^{(2)}{1 \over (2\pi)^3 2 } e^{ip_{x}^{(2)}(x-x_2)}|_{E({\vec p}_{x}^{(2)})}\\
& &N_{x}^{(2)}=({1 \over i\gamma_L^{(2)}})^{1/2}({1 \over i\gamma_T^{(2)}}), \\
& &\gamma_T^{(2)}={1 \over 2}{(t-t_2) \over E({\vec p}_X^{(2)})},\gamma_L^{(2)}=\gamma_T^{(2)}{m^2 \over E({\vec P}_X)^2 } \\
& &{\vec P}_{\vec x}^{(2)}=({\vec x}-{\vec x}_2) {m \over ((t-t_2)^2-({\vec x}-{\vec x}_2)^2)^{1/2}}.   \
\end{eqnarray}
The amplitude becomes, 
\begin{eqnarray}
& & -2i{\lambda}^2 \int d{\vec x}  T_1({\vec X}_l,{\vec P}_l,\cdots,;
{\vec x},t) N_x^{(1)}N_x^{(2)}T_2({\vec X}_m,{\vec P}_m,\cdots,;{\vec x},t), 
\nonumber \\
& & \\
& &T_1({\vec X}_l,{\vec P}_l,\cdots,;{\vec x},t )={1 \over 2(2\pi)^3} \\
& &\times \int d {t_1} d{\vec x}_1 V_g(t_1,{\vec x}_1,{\vec P}_l,{\vec X}_l,T_l) 
{1 \over  E({\vec p}_{x}^{(1)})} e^{ip_{x}^{(1)}(x_1-x)}|_{E({\vec p}_{x}^{(1)})},\nonumber \\
& &  T_2({\vec X}_m,{\vec P}_m,\cdots,;{\vec x},t)^{*}= {1 \over 2(2\pi)^3 } \\
& &\times \int d{t_2} d{\vec x}_2 
 e^{ip_{x}^{(2)}(x-x_2)}|_{E({\vec p}_{x}^{(2)})}
 V_g(t_2,{\vec x}_2,{\vec P}_m,{\vec X}_m,T_m)^{*}.  \nonumber    
\end{eqnarray} 
This amplitude agrees with the previous form if the momentum ${\vec
p}_x^{(1)}$ and ${\vec p}_x^{(2)}$ are regarded as constant vectors.
If these momenta are not  constant vectors, the ${\vec x}$ dependence
and ${\vec x}_j,j=1,2$ dependence of ${\vec p}_x^{(j)},j=1,2$ are taken 
explicitly in the time and coordinate integrations.

\subsection{Factorization}

Factorization is a  general feature of the amplitudes 
obtained in the previous sections. Namely the amplitudes are factorized
into amplitudes of sub-processes that depend on close space and time 
coordinates 
${\vec X}_{i_1},T_{i_1}$ where $|T_{i_1}-T_{j_1}| \approx 0$ and 
${\vec X}_{i_2},T_{i_2}$ where $|T_{i_2}-T_{j_2}|\approx 0$ and
$|T_{i_1}-T_{i_2}| \approx \infty $.  In these short distance
amplitudes, Gaussian momentum integration around minimal are applied 
and amplitudes become almost equivalent to ordinary scattering
amplitudes. On the other hand, the long distance parts have particular
forms that are proportional to the inverses of time difference 
$|T_{i_1}-T_{i_2}|$ in the small 
mass case and to the phases $\exp({im\sqrt{(T_{i_1}-T_{i_2})^2-({\vec
X}_{i_1}-{\vec X}_{i_2})^2}})$ , where $T_{i_1}$ and ${\vec X}_{i_1}$  are  
 average values of times and positions in a group $1$ and $T_{i_2}$ 
and ${\vec X}_{i_2}$ are  average values of times and positions in a 
 group $2$.  The former behavior is due to the expansion of the wave
 packets. It is possible to decompose amplitudes    
in general many body amplitudes in which space time coordinates are
separated into many groups. Each amplitude for the process of close 
coordinates is almost equivalent to ordinary scattering amplitude and
the amplitudes for the long distance parts have the particular 
normalization and the phase factor of the above forms.




\section{Summary}
We have defined the generalized scattering amplitudes which have
dependence upon particle's positions in addition to the particle's 
momenta. Idealistic cases where the positions and the momenta satisfy
minimum uncertainty relations are  studied by the use of minimum wave 
packets, coherent states. 

Since wave
packets are linear combinations of eigenfunctions of free Hamiltonian, 
wave packets change with time.
Wave packets move with a constant group velocity and expand. 
These behaviors occur since each wave of definite 
momentum has a different velocity. They reveal a particle's nature and a
wave nature of wave packets. 
Expansion is slow and has been  irrelevant to any observations in high
energy experiments till recently.  They are relevant in some long
distance experiments and its effects are analyzed in the present work. 
We found also that the expansion speeds satisfy new uncertainty
relations 
expressed in
Eq.(\ref{new-uncertainty1}) and Eq.(\ref{new-uncertainty2}).

Several relations which must be satisfied for the transition 
amplitudes and  probabilities are proven. Completeness of the mixed 
representation is
proven  and is used for defining the weight of phase space integral for 
both variables of momenta and coordinates. The particle states which
are specified by momenta and positions are normalized to unity and Dirac delta
function is unnecessary for the normalization of states in mixed 
representation since the states are normalized but are not extended in
space. The whole  
transition probability from one state to states of a fixed particle
number becomes an intrinsic  value which is independent from the wave
packet sizes and   the total 
transition probability from one state to  
all possible states becomes
unity  under the use of the present measure of
phase space even though states of different momenta and
positions are  nonorthogonal.  So probability interpretation holds. 

Several examples in few body scatterings are analyzed and amplitudes
are explicitly  computed in the lowest order and the second order of
the interaction Hamiltonian.
Translational motions and expansions of wave packets are taken into
account explicitly and their effects are seen in the manifest manner.  It
is shown that scattering amplitudes which have  
long distance part in addition to short distance part are factorized. 
The asymptotic condition for the ordinary scattering, which is
satisfied by the addition of $i\epsilon$ in 
propagators in the standard s-matrix, is realized automatically in the 
present formalism by taking  a suitable limit of the present amplitude.

 Applications to neutrino long distance experiments
 \cite{Asahara},\cite{Ishikawa-shimomura} and others will be
given  in  separate  works.    
     

\section*{Acknowledgements}

This work was partially supported by the special Grant-in-Aid
for Promotion of Education and Science in Hokkaido University
provided by the Ministry of Education, Science, Sports and Culture,
 a Grant-in-Aid for Scientific Research on Priority Area ( Dynamics of
Superstrings and Field Theories, Grant No. 13135201), and a Grant-in-Aid 
for Scientific Research on Priority Area ( Progress in Elementary
Particle Physics of the 21st Century through Discoveries  of Higgs Boson and
Supersymmetry, Grant No. 16081201) provided by 
the Ministry of Education, Science, Sports and Culture, Japan. K.I
thanks T. Ishigaki for useful discussions on measurement problems of
quantum mechanics.   
\\
\\
\appendix
\section{ Generalized vertices of arbitrary  wave packets } 


The product of the wave functions  at $(t,{\vec x})$ are the Gaussian 
function of the space time coordinates $(t,{\vec x})$,
\begin{eqnarray}
& &\Pi_j N_j^{*}e^{-{1 \over 2 \sigma_j}({\vec x}-{\vec X}_j-{\vec v}_j(t-T_j))^2+iE({\vec p}_j)(t-T_j)-i{\vec p}_j({\vec x}-{\vec X}_j)} \nonumber \\
& &\times \Pi_l N_le^{-{1 \over 2 \sigma_l}({\vec x}-{\vec X}_l-{\vec v}_l(t-T_l))^2-iE({\vec p}_l)(t-T_l)+i{\vec p}_l({\vec x}-{\vec X}_l)} \\
& &=\Pi_j N_j^{*} \Pi_l N_l e^{- {1 \over 2 \sigma_s} ({\vec x}-{\vec x}_0(t))^2 - {1 \over 2 \sigma_t} (t-t_0)^2  } e^{R+i\phi}.  \nonumber
\end{eqnarray}
Wave packet parameters in the spatial directions and the temporal
direction are \
\begin{eqnarray}
& &{1 \over  \sigma_s}=\Sigma_j {1 \over  \sigma_j}\\
& &{1 \over  \sigma_t}= \Sigma_j {1 \over  \sigma_j} {\vec v}_j^2-{1 \over \sigma_s}{\vec v}_0^2 
\end{eqnarray}
and the central values of the space-time coordinates are 
\begin{eqnarray}
& & {\vec x}_0(t)={\vec v}_0 t+{\vec x}_0(0), \\
& &{\vec v}_0= \sigma_s \Sigma_j{1 \over \sigma}_j {\vec v}_j,  \\
& &{\vec x}_0(0)=\sigma_s(\Sigma_j{1 \over \sigma_j}{\tilde {\vec X}_j}-i(\Sigma_j (\pm) {\vec p}_j))  \\
& & t_0 = \sigma_t( {1 \over \sigma_s}{\vec v}_0 \cdot {\vec x}_0 -\Sigma_j{1 \over \sigma_j} {\vec v}_j \cdot {\tilde {\vec X}_j}+i\Sigma_j(\pm) E({\vec p}_j)) \\
& &{\tilde {\vec X}_j}={\vec X}_j-{\vec v}_j T_j. 
\end{eqnarray}
The real part determines the magnitude of the amplitude and is composed
of the trajectory terms and the energy-momentum terms. The former give
constraints on the particle trajectories and the latter give constraints 
the total energy and total momentum and are
determined as,
\begin{eqnarray}
& &R=R_{trajectory}+R_{momentum}, \\
& &R_{trajectory}=-\Sigma_j {1 \over 2 \sigma_j}{{\tilde {\vec X}_j}^2+2\sigma_s
({\Sigma_j {1 \over 2 \sigma_j}{\tilde {\vec X}_j} })^2 +2\sigma_t({\Sigma_j ({\vec v}_0-{\vec v}_j){{\tilde {\vec X}_j}}})^2},\nonumber \\
& &  \\
& &R_{momentum}=-{\sigma_t \over 2}({\Sigma_j {(\pm) ({E({\vec p}_j)-{\vec v}_0{\vec p}_j })}})^2 - {\sigma_s \over 2}  (\Sigma_j (\pm) {\vec p}_j)^2.     
\end{eqnarray}
The phase factor is composed of the primary term which expresses the 
energy momentum dependent phase and the secondary terms which are due to 
finite sizes of wave packets and are 
determined as,
\begin{eqnarray}
& &\phi= \phi_0+\phi_1,\\
& &\phi_0= \Sigma_j(\pm)({\vec p}_j{\vec X}_j-E({\vec P}_j)T_j),  \\
& &\phi_1=
-2\sigma_t(\Sigma_j{1 \over 2 \sigma_j}({\vec v}_0-{\vec v}_j){\tilde {\vec X}_j})({{\Sigma (\pm) {\vec v}_0 ({\vec P}_j-E({\vec p}_j) )}})\nonumber \\
& & -2\sigma_s (\Sigma_j(\pm){\vec p}_j )(\Sigma_j{1 \over 2 \sigma_j}{\tilde {\vec X}_j}).
\end{eqnarray}

{}


\begin{thebibliography}{99}
\bibitem{completeness} It should be noted that the following completeness is satisfied for
general wave functions which are localized and have the same phase
factors.  So, generalized S-matrix can be formulated in wide
systems where uncertainty relations between variances of momenta and
coordinates are different from the minimum uncertainty.
\bibitem{classical-phase} In non-relativistic theory,
one particle energy is given as $E({\vec p})={{\vec p}^2 \over 2m}$ and
the stationary phase condition 
${\partial \over \partial p_i} (E({\vec p})t -{\vec p}\cdot{\vec x})=0$
is solved easily. The momentum and the phase at the stationary momentum
are given as  $ m {x_i \over t}, -{m \over 2} ({{\vec x} \over t})^2.$
The phase is not canceled.
\end{thebibliography}
\end{document}